\documentclass{ieeeaccess}

\usepackage{amsmath,amssymb,amsfonts}
\usepackage{graphicx}
\usepackage{textcomp}

\usepackage{lineno,hyperref}
\usepackage[T1]{fontenc}

\usepackage{xcolor}
\usepackage{amssymb}
\usepackage{float}

\usepackage[utf8]{inputenc}
\usepackage [english]{babel}
\usepackage [autostyle, english = american]{csquotes}
\MakeOuterQuote{"}

\usepackage{algorithm}

\usepackage[noend]{algorithmic}
\usepackage{caption}

\newcommand{\TITLE}[1]{\item[#1]}

\newbox\fixbox
\renewcommand{\algorithmicdo}{\setbox\fixbox\hbox{\ {} }\hskip-\wd\fixbox}
\newcommand{\algcost}[2]{\strut\hfill\makebox[1.5cm][l]{#1}\makebox[4cm][l]{#2}}

\newcommand{\nonl}{\renewcommand{\nl}{\let\nl\oldnl}}

\usepackage{textcomp} 
\usepackage{graphicx}
\usepackage{pdfpages}

\usepackage{cite}
\usepackage{amsmath,amssymb,amsfonts}
\usepackage{algorithmic}
\usepackage{graphicx}
\usepackage{textcomp}

\usepackage{amsmath}
\DeclareMathOperator*{\Max}{Max}
\newtheorem{theorem}{Theorem}[section]

\newtheorem{definition}[theorem]{Definition} %

\usepackage{mathtools} 

\def\BibTeX{{\rm B\kern-.05em{\sc i\kern-.025em b}\kern-.08em
    T\kern-.1667em\lower.7ex\hbox{E}\kern-.125emX}}
\begin{document}
\history{Date of publication xxxx 00, 0000, date of current version xxxx 00, 0000.}
\doi{--.----/ACCESS.2021.DOI}

\title{A New Dynamic Optimal M2M RF interface setting in Relay Selection Algorithm (DORSA) for IoT Applications}

\author{\uppercase{Monireh~Allah~Gholi~Ghasri}\authorrefmark{1},
\uppercase{Ali~Mohammad~Afshin~Hemmatyar\authorrefmark{2}}}
\address[1]{Department
of Computer Engineering, Sharif University of Technology, Azadi Ave., Tehran, Iran (e-mail: moghasri@ce.sharif.edu)}
\address[2]{Department
of Computer Engineering, Sharif University of Technology, Azadi Ave., Tehran, Iran (e-mail: hemmatyar@sharif.edu)}


\markboth
{M. A. G. Ghasri \headeretal: A New Dynamic Optimal M2M RF Setting in Relay Selection Algorithm (DORSA) for IoT Apps.}
{M. A. G. Ghasri \headeretal: A New Dynamic Optimal M2M RF Setting in Relay Selection Algorithm (DORSA) for IoT Apps.}

%

\corresp{Corresponding author: Ali~Mohammad~Afshin~Hemmatyar (e-mail: hemmatyar@sharif.edu).}


\begin{abstract}
Machine-to-Machine (M2M) communication is an important type of communication in the Internet-of-Things (IoT). How to send data in these high-density communications using relay selection can help improve the performance of this type of communication in various applications. In addition, the possibility of simultaneous use of different Radio Frequency (RF) interfaces helps to use the spectrum more efficiently. In this work, we try to further use machine communication RF equipment and improve the average data rate of networks in some applications such as the IoT, which have their own bandwidth requirements. Therefore, we provide an optimization algorithm for relay selection as well as the simultaneous and dynamic multiple M2M RF interface setting that is called Dynamic Optimal Relay Selection and RF interfaces Setting Algorithm (DORSA). The simulation results show that the average data rate of DORSA with three interfaces (DORSA\_W-B-Z) can be improved up to 10\% compared to the existing methods such as optimal direct transmission and relay selection algorithms with static RF interface setting.

\end{abstract}


\titlepgskip=-15pt

\maketitle

\section{Introduction}\label{sec:introduction}


Internet-of-Things (IoT) has very diverse applications. Each of these applications has various requirements, such as different required BandWidth (BW), security, or sensitivity to time. For example, energy measuring sensors have lower data BW and less time sensitivity than surveillance and security cameras \cite{bMSAWiLt2012}.

%
Machine-to-Machine communication (M2M) can be considered as a type of communication in which users have the least involvement. These communications are suitable for creating data transfer and decision-making process infrastructure in the world of IoT. In M2M communications, each machine may be equipped with different Radio Frequency (RF) interfaces. These RF interfaces have different communication technology specifications. Some RF interfaces are suitable for low power networks, such as NarrowBand-Internet of Things (NB-IoT) or Z-Wave \cite{sENIPoJh2018}, and some of them are suitable for broadband networks, such as Wireless Fidelity (WiFi) \cite{WiFiwiki2021} or Long Term Evolution (LTE) \cite{LTEtwiki2021}.
Sometimes, IoT applications on machines have data to transmit, and sometimes there is no data for transmission. The machines that have data for transmission are called active machines (sources), and the others are called idle machines. Idle machines can be used as relays in the high-density network \cite{MRSRGhHe2020} to help overcoming bad environmental conditions, such as fading or shadowing in the link between the machines and the Base Station (BS), to increase the network coverage \cite{sCRNAsKh2019}. 


In addition, simultaneous use of different RF interfaces in communications is another solution to increase performance \cite{MRSRGhHe2020, AMHCHuXu2017, PORISiJK2016} and coverage of the network.

Multiple RF interface setting can be adjusted statically or dynamically depending on the source requests, in M2M communications or Machine to Base station (M2B) communications. Other studies usually use one M2M RF interface or static multiple M2M RF interfaces setting \cite{MRSRGhHe2020, AMHCHuXu2017}. In the subject literature, the word dynamic is used as opposed to static. In other words, the term dynamic selection or setting of RF interfaces means the ability to select from multiple RF interfaces in the selection process. Therefore, in this method, the used RF interface is not statically pre-set. In our proposed algorithm, a dynamic optimal M2M RF interface setting with relay selection is provided. In the following, the main contributions of this paper are mentioned in subsection \ref{subsec:contribution}.

\subsection{Main Contributions}\label{subsec:contribution}

In this paper, a novel optimal M2M RF interface setting and relay selection algorithm is proposed. This algorithm has the following main contributions: 

\begin{enumerate}

\item[-] This is an optimal algorithm for simultanously M2M RF interface setting along with relay selection for M2M communication with multiple M2M RF interfaces by transforming this problem to a $k$-AP. 

\item[-] This proposed RF interface setting method for M2M communications is done dynamically according to the amount of source requests. This proposed algorithm is suitable for situations where IoT applications with similar BW requests are running on network machines.


\end{enumerate}
\begin{enumerate}
\item[•] This paper can be considered as the first work in the field of dynamic M2M RF interface setting, which simultaneously selects the dynamic optimal relay-M2M RF interface pair to maximize the average network data rate.
\end{enumerate}

The rest of this paper is organized as follows. Section \ref{sec:transJRRSS} provides the background of problem transformation. Some related works are reviewed in section \ref{sec:relWork}. The system model of our algorithm is provided in section \ref{sec:sysModel}. Then, section \ref{sec:proDORSA} describes the proposed dynamic optimal RF interface setting in the relay selection algorithm. In the following, the results of simulations are studied in section \ref{sec:simulation}. Finally, section \ref{sec:conclusion} provides the conclusions of this paper.

\section{Background of Problem Transformation}\label{sec:transJRRSS}

\begin{figure}[htb]
\centering
\includegraphics[scale=0.5]{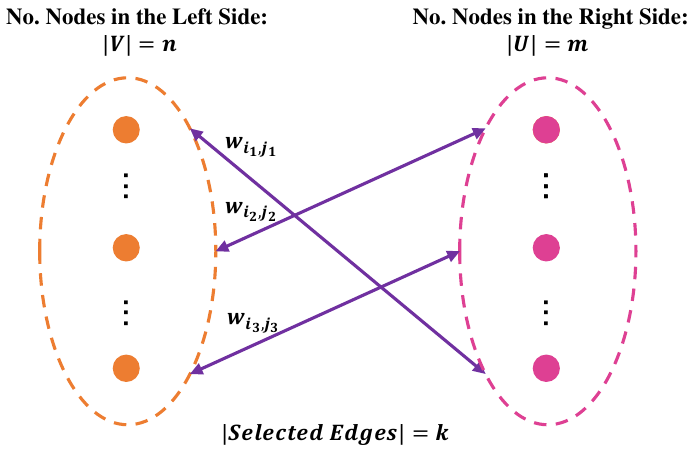}
\caption[The bipartite graph of the $k$-cardinality assignment problem.]{The bipartite graph of the $k$-cardinality assignment problem.}
\label{fig:kAPRef}
\end{figure}

Before starting to solve the problem, we will explain the concepts that are used below. Since graphic models can well model network problems, to solve the desired joint relay selection and dynamic RF interface setting problem, which is formulated in equation~(\ref{eq:optEq_srRFb}), it can be expressed as a maximum weighted matching in a bipartite graph. Therefore, it can be transformed into a problem that can be solved with mathematical tools. For this purpose, in the proposed algorithm (section \ref{sec:proDORSA}), we first model the problem as a graph and then consider that problem as a kind of maximum weighted matching problem with $k$-edge constraint ($k$-cardinality assignment problem ($k$-AP)). Next, we solve the new problem by transforming it into a standard assignment problem using existing mathematical tools. Therefore, we introduce this mathematical tool in this section and then, details of our proposed algorithm is provided in section \ref{sec:proDORSA}. A schematic of the bipartite graph of the $k$-AP is shown in Figure~\ref{fig:kAPRef}.


 This model of bipartite weighted graph is known as $BG=(V \cup U, E)$ where:
\begin{itemize}
\item[-] the set of vertices is called by $ \lbrace V \cup U \rbrace $,
\item[-] $|V|=n$ and $|U|=m$,
\item[-] the set of edges is called by $E=\lbrace (v_i, u_j)| v_i \in V  \wedge  u_j \in U \rbrace $, and
\item[-] the cost of edge $(v_i, u_j)$ is  $w_(i, j)$.
\end{itemize}

Now, the $k$-AP and related concepts are defined as follows:

\begin{definition}
\label{def:kAP}
\textbf{k-cardinality Assignment Problem(k-AP):} An assignment problem that selects $k$ edges, where $(k \leq min\lbrace m, n \rbrace )$, in a bipartite graph so that the total weight of the selected edges is maximized, the problem is known as a $k$-cardinality problem.
\end{definition}

A $k$-AP can be solved with a polynomial solver \cite{DJThMARC2015} but we know that it can not be solved directly using the Hungarian algorithm. Therefore, to make it easier to solve in the following steps:

\begin{definition}
\label{def:orsaSteps}
\textbf{Steps to solve the problem is modeled as a $k$-AP with Hungarian algorithm \cite{MRSRGhHe2020}:} 
 
 \begin{itemize}
\item[•] Step 1: Transforming our problem is modeled as a $k$-AP to a standard assignment problem by adding additional nodes,
\item[•] Step 2: Solving the new standard assignment problem with Hungarian algorithm and obtaining final results of our problem.
 \end{itemize}

\end{definition}

In the following, we define the standard assignment problem:

\begin{definition}
\label{def:standardAP}
\textbf{Standard Assignment Problem:} A $k$-AP without any constraint on the number of selected edges. In other words, in this problem $k=n$ or $k=m$. Therefore, it can be said that the $k$-cardinality assignment problem is a generalization of the standard assignment problem  \cite{skctVolg2004, kcapDeMa1997}. 
\end{definition}

Some mathematical tools are used to solve these assignment problems. One of these solver tools is the Hungarian Algorithm.

\begin{definition}
\label{def:HungAlg}
\textbf{Hungarian Algorithm:} This is a common solver to obtain optimal solution for a standard assignment problem as a maximum weighted matching problem \cite{ICRDchitr2016, KuhaHMAP1955}. This algorithm can solve a standard assignment problem with a polynomial-time complexity is equal to $O(n^3)$ where $n$ is the number of vertices of this problem\cite{DHACMiSt2007, GithFCIH2019}.
\end{definition}


%
%
%

In other words, firstly, we can transform the $k$-AP into a standard assignment problem and then solve it using the Hungarian algorithm \cite{MRSRGhHe2020}. This solution method has already been used to solve the relay selection problem with static RF interfaces setting and provide the Optimal Relay Selection Algorithm (ORSA) \cite{MRSRGhHe2020}.

In section \ref{sec:proDORSA}, we will describe details of our proposed algorithms include the steps of transforming desired joint relay selection and dynamic RF interfaces setting problem to the $k$-AP, solving the transformed problem, and then obtaining the final results in three steps.


\section{Related Works}\label{sec:relWork}

In the related works section, a review of some works about relay selection and RF interface setting problems is summarized.



\subsubsection{Relay Selection }

%
The next-hop when sending data can be selected according to the channel conditions and data rate between the data source and the desired relay \cite{sCRNAsKh2019}. Selecting an appropriate relay can help to send data in situations where direct transmission has low quality \cite{RR5HDaPa2018, CRRMBaLo2011}. In this subsection, we give a brief overview of the works done in this area.




There are lots of works in the field of relay selection in different networks. As mentioned earlier, in this paper we seek to solve the problem of optimal selecting and assigning RF interfaces and relays simultaneously. Finally, by transforming this problem into a standard assignment problem, we use the Hungarian algorithm as a solver to find desired maximum weighted matching. Therefore, in this subsection, we focus on reviewing some relay selection algorithms that are designed using the Hungarian algorithm as a mathematical tool that can be used as a solver for the maximum weighted matching problems in bipartite weighted graphs \cite{ICRDchitr2016}.





\begin{itemize}
\item[-] Some works in \textbf{Relay Selection using by Hungarian Algorithm} :
\end{itemize}


In a recent study, two new relay selection algorithms for M2M communication with static RF interfaces setting were provided \cite{MRSRGhHe2020}. An Optimal Relay Selection Algorithm (ORSA) was proposed by converting this problem to a K-cardinality assignment problem. Then, the problem was solved by the Hungarian algorithm. Furthermore, a Matching based Relay Selection Algorithm {MRSA} was provided by deferred acceptance procedure. The result of MRSA is a stable matching \cite{MRSRGhHe2020}. In section \ref{sec:simulation}, these algorithms were compared with our new algorithms in this paper.



In another study, an NB-IoT environment was considered \cite{ENMRChHu2018}. In this study, in addition to maintaining the throughput and Quality of Service (QoS) in the system, which is formed by appropriate selection of relays in M2M communications with NB-IoT interface, the need for repetitive data transmission and energy consumption are also reduced. Therefore, the problem was modeled as a weighted bipartite matching problem and then was solved using the Hungarian algorithm \cite{ENMRChHu2018}.

In other work, an iterative Hungarian method (IHM) was proposed in Device to Device (D2D) communications \cite{IHRRKiDo2014}. This method solves the relay selection and resource allocation problem by achieving a near-optimal solution \cite{IHRRKiDo2014}.

In another study, the problem of relay selection and channel allocation in a cognitive network transforms to a classical weighted bipartite graph matching problem. Then, the new problem was solved with the aim of maximization Signal-to-Interference-plus-Noise-Ratio (SINR) using the Hungarian algorithm \cite{GRSCAlFo2011}.		

\subsubsection{Multiple RF Interfaces Setting}



Network devices can use multiple RF interfaces in their communications. Selecting and using any RF interface on devices requires configuration. For example, in default handover settings from one RF interface to another on smartphones are done automatically or manually by the user. In this regard, some papers offer different criteria for increasing the performance of automatic handover on smartphones \cite{WLHMBerg2019, LWLSFoGr2015, WSTBGrFo2015}.

On the other hand, simultaneous use of multiple RF interfaces can help increase the average network data rate, for example by distributing traffic on different RF interfaces \cite{ORODWuHe2018, RAMDElCh2015}. In addition, this simultaneous use can be effective in reducing interference due to the simultaneous transmission of data on an RF interface. 

In this regard, one of the recent works by presenting a method based on a fuzzy system made it possible to connect several RF interfaces simultaneously during handover and change of access point \cite{HMSFCaSi2021}. In another recent work, as mentioned, the static simultaneously setting of two M2M RF interfaces was used when selecting the next hop \cite{MRSRGhHe2020}. The simultaneous use of predefined communication interfaces focused on limited feedback for multicasting is presented in  another paper\cite{AMHCHuXu2017}. Finally, in another study, while providing a parallel routing method, the simultaneous use of multiple RF interfaces when sending data is mentioned \cite{PORISiJK2016}

\section{System Model}\label{sec:sysModel}

In this section, the system model of our algorithm is described. Uplink paths in a cell with one BS and $N$ fixed machines are considered for this model. The BS is located in the middle of the cell and the machines are randomly placed around it with a uniform distribution. Each machine is equipped with multiple RF interfaces for M2M communications (M2M RF interfaces) and one RF interface for M2B communications (M2B RF interface). In these communications, machines can use the network infrastructure (the BS) for data transmission and can perform M2M RF interface setting without any user intervention.

Machines are divided into two following sets in the desired time snapshot:
\begin{itemize}
\item[-] Active machines: machines that have data to send to their destinations by BS and They are also called sources,
\item[-] Idle machines: machines that have no data to send which are hereinafter referred to as relays. 
\end{itemize}  

It is assumed these machines cooperate to increase the average network data rate. In other words, when the conditions for direct connection of sources to the BS are not favorable, the idle machines can help send data of the sources. Therefore, They act as relays.

The set of sources and relays are denoted by $S$ and $R$, respectively, and each of them includes $N^s = | S |$ and $N^r = | R |$ machines. Therefore, the set of all machines inside the cell can be denoted by $|M|$, such that $M= S \cup R$, $S  \cap R= \phi$, and $|M| = N = N^s + N^r$. The following are the assumptions related to the desired problem.

\subsection{Problem Assumptions}\label{sec:probAssumptions}

In the following, a summary of the system model assumptions is given:
\begin{enumerate}

\item \label{ass:no1}All sources have the similar requested BW.
\item \label{ass:no2}Different RF interfaces frequencies (WiFi, Bluetooth, Z-Wave, and LTE) are not overlapping with each other and they do not interfere.
\item \label{ass:no4}The relationship between sources with relay-RF interface pairs is one-to-one
%
%

\item \label{ass:no6}The Decode-and-Forward (DF) protocol is used as relaying protocol.
\item \label{ass:no8}Due to the placement, the probability of having Line-of-Sight (LoS) propagation between network devices is low, and as a result, we assumed that we do not have Line-of-sight (LoS) propagation in communication.
\end{enumerate}

In the next subsection, problem formulation and its background are provided.
%





\subsection{Problem Formulation}

\begin{figure}[!htb]
\centering
\includegraphics[scale=0.40]{{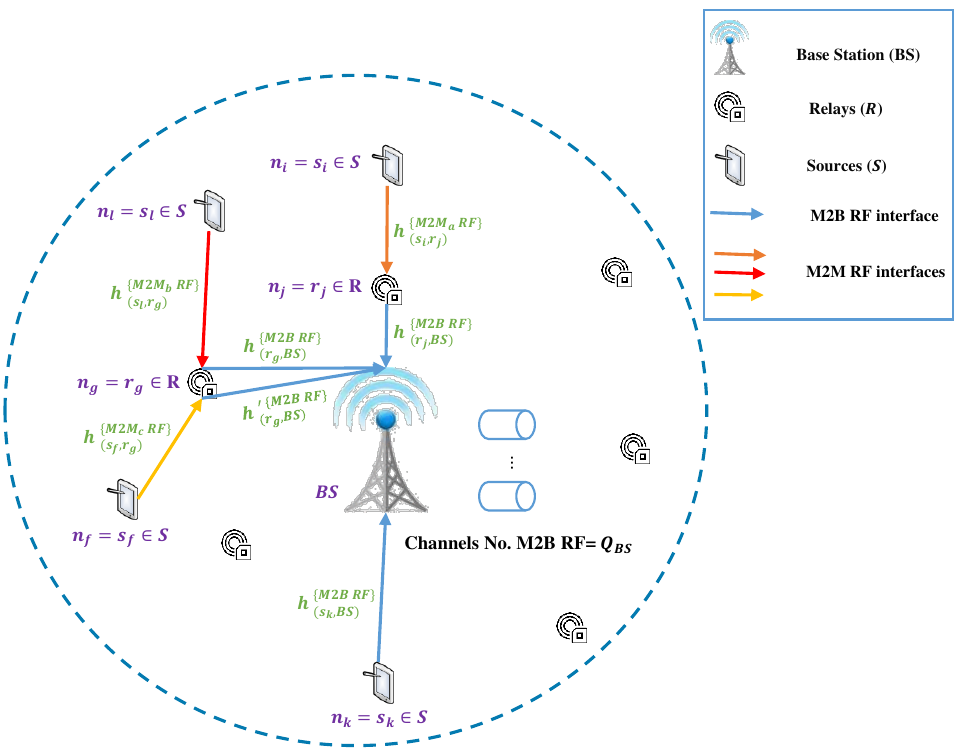}}
\caption[The schematic of sources and relays in the cell of the system model.]{The schematic of sources and relays in the cell of the system model.}
\label{fig:myScenario}
\end{figure}

Figure \ref{fig:myScenario} shows the schematic of sources, relays, and BS communications in the cell of the system model. 

In this figure (Figure \ref{fig:myScenario}), $h^{ \lbrace M2M_{a} RF \rbrace }_{(s_i, n_j)}$ and $h^{ \lbrace M2B RF \rbrace }_{(n_i, BS)}$  are the gains of the channel between the $i$th and $j$th nodes on $a$th M2M RF interface and between the $i$th node and BS on $M2B$ RF interface, respectively. These gains are modeled with sum of path loss, shadowing, and small scale fading on $a$th $M2M$ or $M2B$ RF interface channel between the $i$th node and $j$th node or BS, as presented by equation~(\ref{eq:channelgainM2M}) and equation~(\ref{eq:channelgainM2B}). 

\begin{align}
\label{eq:channelgainM2M}
    & h^{ \lbrace M2M_{a} RF \rbrace }_{(n_i, n_j)}(dB) 
    \begin{aligned}[t]
      &= Path Loss_{(n_i, n_j)}(dB) \\
       &+ Shadowing^{ \lbrace M2M_{a} RF \rbrace } (dB) \\
       & + Small  Scale  Fading^{ \lbrace M2M_{a} RF \rbrace } (dB),  \\
    \end{aligned} \notag \\
\end{align}

\begin{align}
\label{eq:channelgainM2B}
    & h^{ \lbrace M2B RF \rbrace }_{(n_i, BS)}(dB) 
    \begin{aligned}[t]
      &= Path Loss_{(n_i, BS)}(dB) \\
       &+ Shadowing^{ \lbrace M2B RF \rbrace } (dB) \\
       & + Small  Scale  Fading^{ \lbrace M2B RF \rbrace } (dB),  \\
    \end{aligned} \notag \\
\end{align}
 
where:
\begin{itemize}
\item[-] Shadowing (dB) on $a$th $M2M$ or $M2B$ RF interface is modeled by a Normal random variable with zero mean and standard deviation of 8 ($\mathcal{N}(0,64)$), 
\item[-] Small scale fading (dB) is modeled by a Rayleigh random variable with scale parameter of $\sigma_r = 1$ (according to assumption \ref{ass:no8}), and
\item[-] Path loss between $i$th node and $j$ node or BS is modeled with path loss exponent $\beta=4$ (equation~(\ref{eq:pathloss4})).
\end{itemize}

\begin{equation}
\label{eq:pathloss4}
Path Loss_{(n_i, (n_j | BS))}(dB)=10 \beta log_{10}(\frac{d_{(n_i, (n_j | BS))}}{d_{0}},
\end{equation}

where $d_{0}=10 (m)$, and $d_{(i,j)}$ is the euclidean distance between node $i$ and node $j$.

The maximum channel data rate capacity is calculated using the Shannon-Hartley theorem. These maximum data rates between the $i$th and $j$th nodes on $a$th M2M RF interface and between the $i$th node and BS on $M2B$ RF interface, respectively, is provided by equation~(\ref{eq:Capninj}) and equation~(\ref{eq:CapniBS}).

\begin{equation}
\label{eq:Capninj}
	C^{ \lbrace M2M_{a} RF \rbrace }_{(n_i, n_j)} = B^{ \lbrace M2M_{a} RF \rbrace }_{(n_i, n_j)}   log_2(1+\mathrm{SINR^{ \lbrace M2M_{a} RF \rbrace }_{(n_i, n_j)}}),
\end{equation}	

\begin{equation}
\label{eq:CapniBS}
	C^{ \lbrace M2B RF \rbrace }_{(n_i, BS)} = B^{ \lbrace M2B RF \rbrace }_{(n_i, BS)}   log_2(1+\mathrm{SINR^{ \lbrace M2B RF \rbrace }_{(n_i, BS)}}),
\end{equation}	

where:
\begin{itemize}
\item[-]  $B^{ \lbrace M2M_{a} RF \rbrace }_{(n_i, n_j)}$ and $B^{ \lbrace M2B RF \rbrace }_{(n_i, BS)}$ are the bandwidth of the channel between the $i$th and $j$th nodes on $a$th M2M RF interface and between the $i$th node and BS on $M2B$ RF interface, respectively,

\item[-]  $SINR^{ \lbrace M2M_{a} RF \rbrace }_{(n_i, n_j)}$ and $SINR^{ \lbrace M2B RF \rbrace }_{(n_i, BS)}$ are the achieved SINR of the channel on $a$th $M2M$ and $M2B$ RF interface between the $i$th and $j$th nodes and between the $i$th node and BS, respectively. These are computed by equation~(\ref{eq:sinrM2M}) and equation~(\ref{eq:sinrM2B}).

\end{itemize}

\begin{equation}
\label{eq:sinrM2M}
	\mathrm{SINR^{ \lbrace M2M_{a} RF \rbrace }_{(n_i, n_j)}} = \frac{P_{n_i}^{\mathrm{ \lbrace M2M_{a} RF \rbrace}} \times h_{(n_i, n_j)} }{\sigma^2 + \sum_{(k\in S, k \neq i)} P_{n_k}^{\mathrm{ \lbrace M2M_{a} RF \rbrace}} \times h_{(n_k, n_j)} },
\end{equation}	
	
\begin{equation}
\label{eq:sinrM2B}
	\mathrm{SINR}_{(n_i, BS)}^{ \lbrace \mathrm{M2B RF} \rbrace } = \frac{P_{n_i}^{\mathrm{ \lbrace M2B RF \rbrace}} \times h_{(n_i, BS)} }{\sigma^2 },
\end{equation}

where $P_{n_i}^{ \lbrace M2M_{a} RF \rbrace }$ and $P_{n_i}^{ \lbrace M2B RF \rbrace }$ are the transmission powers of the channel on $a$th $M2M$ and $M2B$ RF interface between the $i$th and $j$th nodes and between the $i$th node and BS, respectively. $\sigma^2$ is noise power and $h^{ \lbrace M2B RF \rbrace }_{(n_i, BS)}$  are the gains of the channel on $a$th $M2M$ and $M2B$ RF interface between the $i$th and $j$th nodes and between the $i$th node and BS, respectively, are formulated by equation~(\ref{eq:channelgainM2M}) and equation~(\ref{eq:channelgainM2B}). 

In our system model, if a source transmits its data to the BS in one hop, the data rate between them is calculated according to equation~(\ref{eq:CapniBS}). But if the source $s$ sends its data to the BS in two-hop through a relay $r$ based on DF relaying in a time slot \cite{IRSTZhXu2013}, its data rate is calculated by equation~(\ref{eq:Capsrd}).

\begin{equation}
\label{eq:Capsrd}
	C_{{s,BS}_{in two hop}} = min \lbrace C^{ \lbrace M2M_{a} RF \rbrace }_{s,r}, C^{ \lbrace M2B RF \rbrace }_{r,BS} \rbrace,
\end{equation}

where $C^{ \lbrace M2M_{a} RF \rbrace }_{s,r}$ is the maximum data rate between the source and the relay on $a$th $M2M$ and $M2B$ RF interface and $C^{ \lbrace M2B RF \rbrace }_{r,BS}$ is the maximum data rate between the source and the BS on $M2B$ RF interface.

In the next subsection, the desired problem will be formulated.

\subsubsection{\textbf{Optimization Problem Formulation}}
\begin{figure}[!htb]
\centering
\includegraphics[scale=0.38]{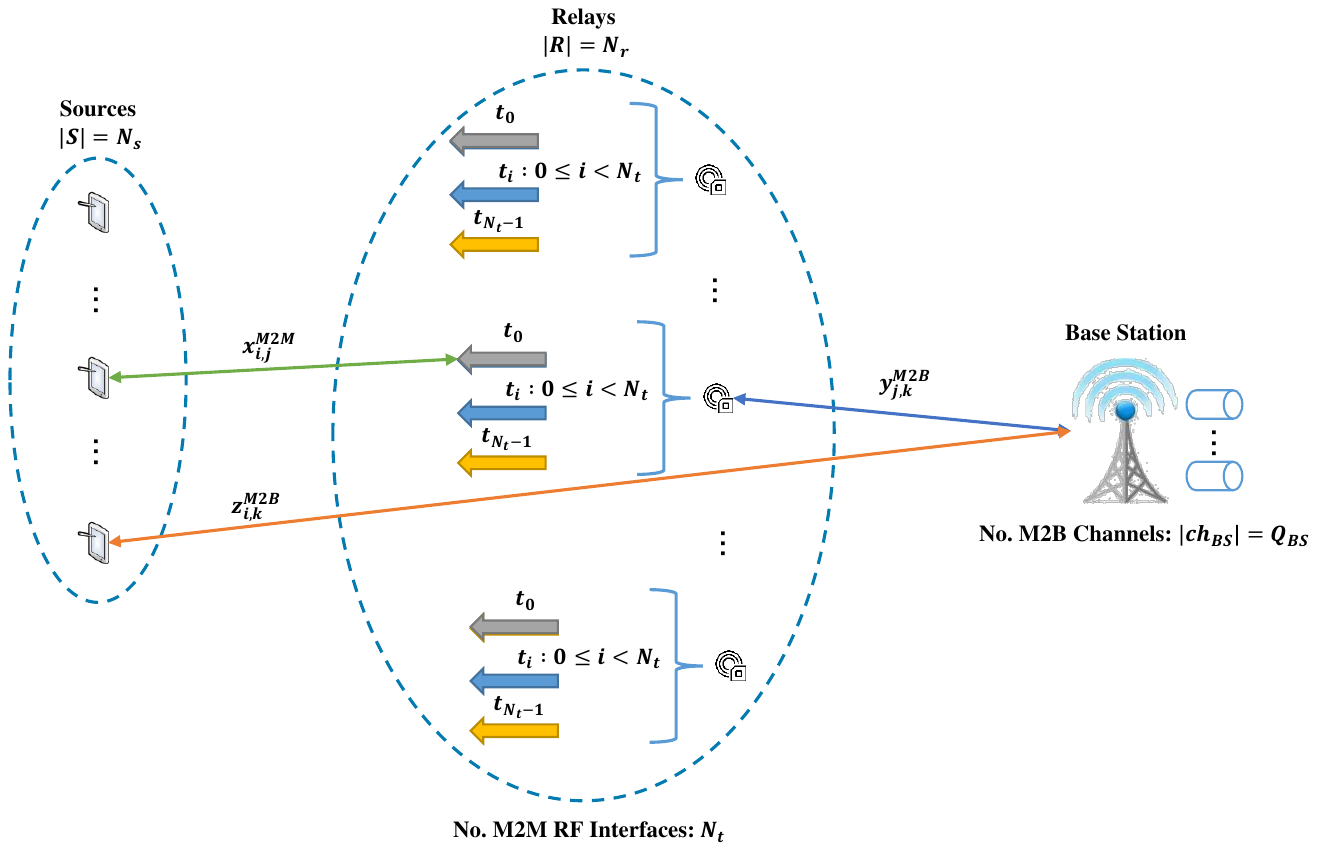}
\caption[The schematic of the graph model of our dynamic RF interfaces setting and relay selection problem.]{The schematic of the graph model of our dynamic RF interfaces setting and relay selection problem.}
\label{fig:selectionModel}
\end{figure}

Here, the optimized relay selection and dynamic RF interfaces setting are formulated by equation~(\ref{eq:optEq_srRFb}). The schematic of the graph model of our dynamic RF interfaces setting and relay selection problem is shown in Figure~\ref{fig:selectionModel}.

\begin{align}
\label{eq:optEq_srRFb}
    &\Max_{ x, y, z}
    \begin{aligned}[t]
       &\sum_{i=0}^{(N_s-1)}{\sum_{j=0}^{(N_t N_r-1)}\sum_{k=0}^{0}  x_{i,j} y_{j,k} \times min( c_{i,j}, c^{''}_{j,k} )} \\
       & +  		\sum_{i=0}^{(N_s-1)}\sum_{k=0}^{0} z_{i,k} c^{'}_{i,k},  \\
    \end{aligned} \notag \\\\  
        \text{Subject to} \notag \\
    & x_{i,j} \in \lbrace 0, 1 \rbrace: \quad for \quad 0 \leq i < N_s , \label{eq:optEq_xij} \\   
    & y_{j,k} \in \lbrace 0, 1 \rbrace: \quad for \quad 0 \leq j < N_t N_r, \label{eq:optEq_yij}\\   
    & z_{i,k} \in \lbrace 0, 1 \rbrace: \quad for \quad k=0,  \label{eq:optEq_zij}\\   
    & \sum_{i=0}^{(N_s-1)} x_{i,j} \leq 1 , \sum_{k=0}^{0} y_{j,k} \leq 1 : for (0 \leq j < N_t N_r), \label{eq:optEq_firstnEq}\\
    & \sum_{j=0}^{(N_t N_r-1)} x_{i,j} \leq 1 , \sum_{k=0}^{0} z_{i,k} \leq 1 : for (0 \leq i < N_s), \label{eq:optEq_secnEq}\\    
& \sum_{i=0}^{(N_s-1)}{\sum_{j=0}^{(N_t N_r-1)}\sum_{k=0}^{0}  x_{i,j} y_{j,k}} \notag \\
& + \sum_{i=0}^{(N_s-1)}\sum_{k=0}^{0} z_{i,k} \leq Q_{BS} \label{eq:optEq_forthnEq}.
\end{align}


The definitions of the used variables are as follows: 

\begin{itemize}
\item[-] $x_{i,j}$: is $1$ if $i$th source has selected the $j$th relay-M2M RF interface pair and $0$ otherwise,
\item[-] $y_{j,k}$: is $1$ if $j$th relay-M2M RF interface pair has selected the $k$th base station and $0$ otherwise,
\item[-] $z_{i,k}$ : is $1$ if $i$th source has selected the $k$th base station and $0$ otherwise,
\item[-] $c_{i,j}$: the capacity between $i$th source and $j$th relay-M2M RF interface pair,
\item[-] $ c^{'}_{i,k}$: the capacity between $i$th source and $k$th base station,
\item[-] $c^{''}_{j,k}$: the capacity between $j$th relay-M2M RF interface pair and $k$th base station, 
\item[-] $N_s$: the number of sources,
\item[-] $N_r$: the number of relays,
\item[-] $N_t$: the number of M2M RF interfaces,

\item[-] $Q_{BS}$: quota or connection capacity of the BS. This is equivalent to the maximum available number of $M2B$ RF interface channels, which is calculated by the following equation:

\begin{equation}
\label{eq:QBS}
	Q_{BS} = \frac{BW^{\mathrm{M2B RF}}}{BW_{MaxReq}},
\end{equation}

where $BW^{M2B}$ is the BW of M2B RF interface for BS and $BW_{MaxReq}$ is the maximum requested BW of sources. While as mentioned in subsection \ref{sec:probAssumptions}, this model assumes that the requested BW of all sources is equal. Hereinafter,  $Q_{BS}$ is referred to as the number of M2B channels.

%
%

\item[-] The constraint is that each relay-M2M RF interface pair can only be assigned to a single source is represented by the first summation in equation~(\ref{eq:optEq_firstnEq}).

\item[-] The constraint is that each relay-M2M RF interface pair can only be connected to a single BS is represented by the second summation in equation~(\ref{eq:optEq_firstnEq}). (Although in our model only one base station is considered, this condition is written in general.)

\item[-] The constraint is that each source can only be connected to a single relay-M2M RF interface pair is represented by the first summation in equation~(\ref{eq:optEq_secnEq}).

\item[-] The constraint that each source can only be connected to a single BS is represented by the second summation in equation~(\ref{eq:optEq_secnEq}). (Although in our model only one base station is considered, this condition is written in general.)

\item[-] The first summation in equation~(\ref{eq:optEq_forthnEq}) represents the total number of two-hop connections of sources to BS through relays on $a$th M2M RF interface and the second summation in equation~(\ref{eq:optEq_forthnEq}) represents the total number of direct connections of sources to the base station on $M2B$ RF interface. 
The summation of the total number of two-hop connections of sources to BS through relays and the total number of direct connections of sources to the base station is less than or equal to the maximum available number of $M2B$ RF interface channels for connection to the base station ($Q_BS$).

\end{itemize}

\section{Proposed Centralized Solution for the Joint Relay Selection and Dynamic RF Interfaces Setting problem}\label{sec:proDORSA}

%
%

%
In this section, details of the proposed optimized solution for the joint dynamic RF interfaces setting and relay selection problem in M2M communications for similar IoT applications are described. Paying attention to how to optimally select the simultaneous RF interface along with selecting the relay in our proposed algorithm is one of the first researches in this field. To solve this problem, we transform it into a k-cardinality Assignment Problem($k$-AP). It has already been proven that a $k$-AP has a polynomial solver \cite{DJThMARC2015}. Finally, the transformation process will be explained.

\subsection{\textbf{Step 0: Transforming Joint Relay Selection and Dynamic RF Interfaces Setting Problem to a k-AP}}\label{subsec:step0transJRRSS}


The relay selection and dynamic RF interfaces setting problem can be solved by transforming it into a $k$-AP. Then, the transformed problem is solved with a solver that has already been used to design the Optimal Relay Selection Algorithm (ORSA) with static RF interfaces setting \cite{MRSRGhHe2020}. 

Now, the bipartite weighted graph corresponding to the problem statement in equation~(\ref{eq:optEq_firstnEq}) (shown in Figure~\ref{fig:selectionModel}), should be defined. In other words, all nodes in the network should be divided into two parts with one-to-one relation in its graph model. Therefore, we need to define the vertices of the two parts of the bipartite graph, so that the relationship between the vertices of the two parts is at most one to one. For this purpose, as in the case of selecting a relay by static RF interfaces setting \cite{MRSRGhHe2020}, we can place the sources in one part, for example on the left side, and the relays and base station in another part, for example on the right side.

But here, given that each relay with each of its M2M RF interfaces can connect to a maximum of one source, according to the amount of source requests, (assumption \ref{ass:no4}), then connect to the BS to send data of all connected sources. Thus, if the model is exactly the same as before, the relationship between the relays and the sources will be one-to-many. Then, to better transformation modeling, each relay with one of its M2M RF interfaces is considered an entity (a vertex of the right side in the bipartite weighted graph). 


Therefore, we have a bipartite weighted graph $G=(V \cup U, E)$, where set of vertices is $ \lbrace V \cup U \rbrace $, with $n = N_s$ nodes in the left side and $m = N_r \times  N_t + Q_{BS} $ nodes in the right side. The weights of the edges in this weighted bipartite graph are equal to the data rate of the path corresponding to the two nodes connected to that edge according to equation~(\ref{eq:CapniBS}) in direct path ($= c_{i_1,j_1}$) and equation~(\ref{eq:Capsrd}) in two hop path ($= c_{i_2,j_2}$). Finally, $k = Q_{BS}$ edge is to be selected from all the edges in this graph so that the total weight of these edges is maximized.

\begin{figure}[htb]
\centering
\includegraphics[scale=0.5]{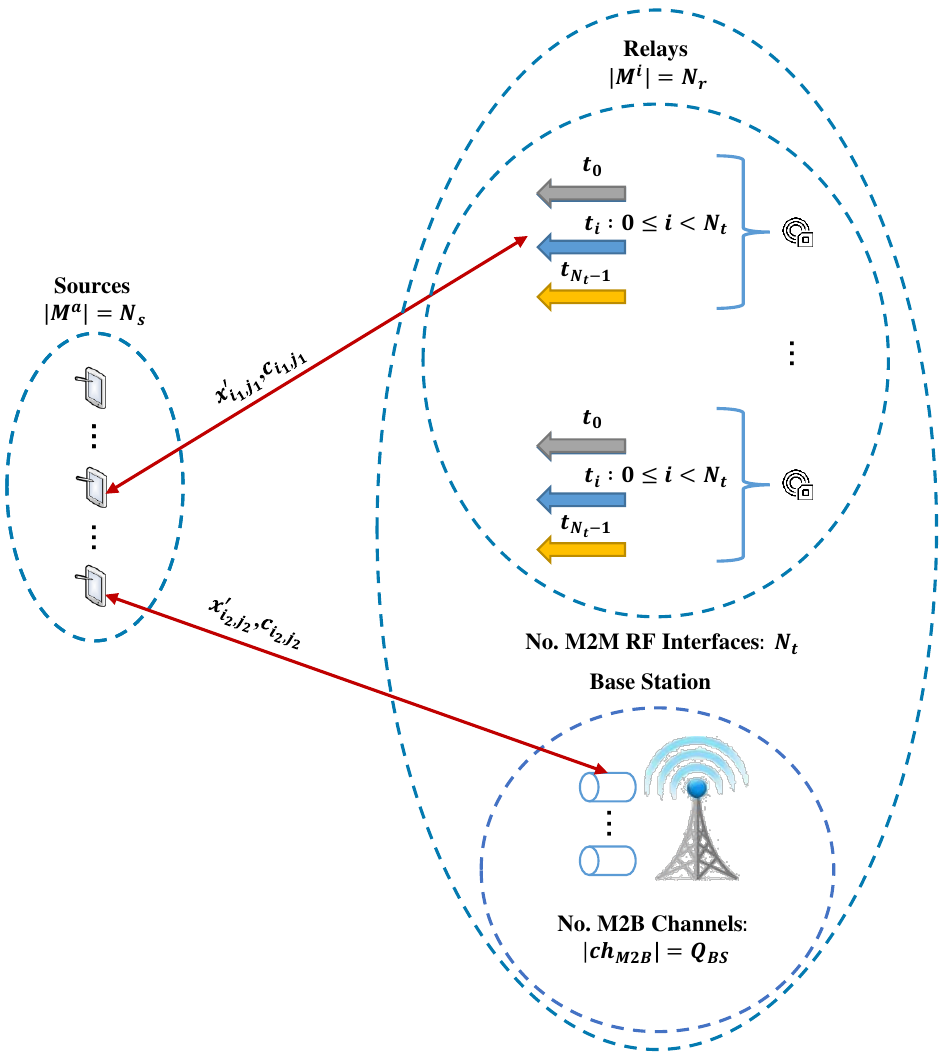}
\caption[The bipartite graph of the joint relay selection and dynamic RF interfaces setting problem.]{The bipartite graph of the joint relay selection and dynamic RF interfaces setting problem.}
\label{fig:InitialRSABiparGraph}
\end{figure}

Figure~\ref{fig:InitialRSABiparGraph} shows a schematic of the bipartite graph of the joint relay selection and dynamic RF interfaces setting problem as described.

\subsection{\textbf{Step 1: Transforming transformed k-AP to a standard assignment problem and solve it}}\label{subsec:solveJRRSS}

%
%
%
%

After transforming our problem to the $k$-AP in the previous step, we should look for a solution to the new transformed problem ($k$-AP). Since the Hungarian algorithm is a convenient tool to solve a standard assignment problem, such as step 1 in definition \ref{def:orsaSteps}, we can transform the $k$-AP to an equivalent standard assignment problem. 

We know that in the main problem if $n=N_s$ and $m=N_r \times  N_t + Q_{BS}$, after the final matching, the maximum $k=Q_{BS}$ sources can be connected to one of the nodes on the right. Therefore, in the final match at least $n-k = N_s-Q_{BS}$ sources, $m-k=(N_rN_t + Q_{BS}) - Q_{BS})$ relay-RF interface pair and base station channels will remain without connection.

 
In other words, our ultimate goal in solving this problem is to determine which of the $Q_{BS}$ sources are connected by which relays or directly to the base station so that the total data rate of the final connections is maximized. To have a standard problem, the limitation of the number of edges must be solved without changing the principle of the problem. So, without losing the generality of the problem, we can add the additional nodes to both sides. Thus, if we can match the sources and relays that are not the maximum matching with the new nodes (as new vertices in the graph), any connection in the final match that is established between the previous nodes (sources on the right side and the relays and channels of the base station) is our answer.


%

Now without losing the generality of the problem, we add the $n-k =N_s-Q_{BS}$ node to the right side of the graph and the $m-k=N_rN_t$ node to the left side of the graph. Thus the number of vertices in both parts of the graph will be equal to $N_rN_t + N_s$, and the final one-to-one matching between the two parts of the new graph will be a standard matching problem or $k_{new}$-AP with $k_{new}=N_rN_t+N_s$. 




Thus, if the new nodes of each part are connected:
\begin{itemize}
\item[-] to the previous nodes of the opposite part with an edge of infinite weight (or in practice equal to a large enough value ($A_{value}$) such as $1 + \sum_{e \in E} w_e$ ) and
\item[-] to the new nodes of the opposite part with an edge of infinite weight or zero (which we then considered zero),
\end{itemize}
 the final matching includes:
 \begin{itemize}
\item[-] the maximum $Q_{BS}$ edge of The previous sources and nodes are to the right of the graph (relay-interface or base station) and
\item[-] the edges are connected to the new nodes with nodes on the opposite side (previous nodes with $A_{value}$ weight and new nodes with zero).
\end{itemize}


{Note: Due to the edges between the new nodes with each other are not examined in the final matching, their weight will be ineffective. In the following, the weight between the new nodes on both parts is considered equal to zero, which will not affect the total weight of the edges. Therefore, the generality of the problem does not disappear.}



\begin{itemize}
\item[•] A point to considered: For less time complexity, if the number of sources ($N_s$) is less than the number of M2B channels ($Q_{BS}$), it is better to replace the number of M2B channels with the number of sources ($Q_{BS}=N_s$). Thus the number of nodes in the graph that directly affect the execution time of the algorithm can be reduced.
\end{itemize}

\begin{figure}[htb]
\centering
\includegraphics[scale=0.5]{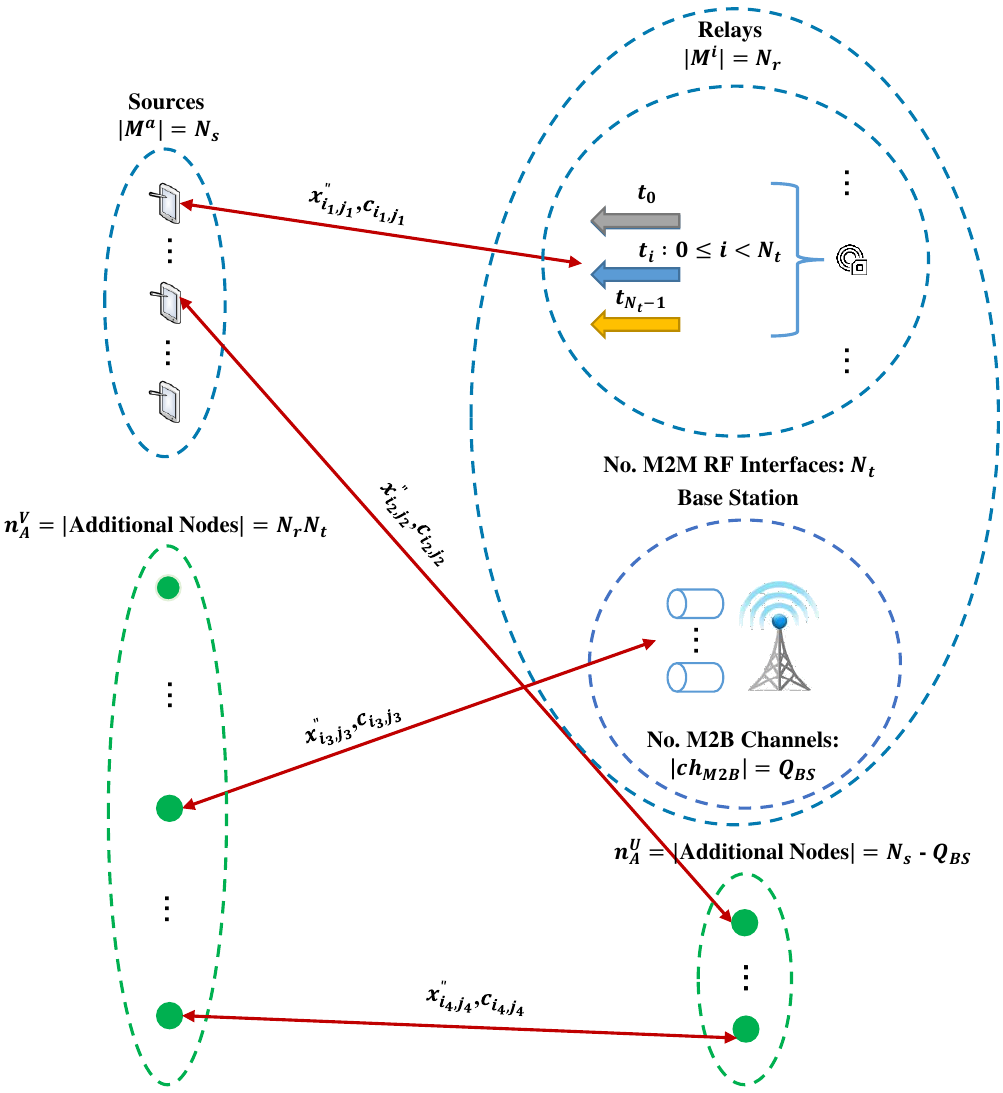}
\caption[The bipartite graph of the joint relay selection and dynamic RF interfaces setting problem.]{The bipartite graph of the joint relay selection and dynamic RF interfaces setting problem.}
\label{fig:kAPRSAGraph}
\end{figure}

The transformed bipartite graph model of the joint dynamic relay-RF interfaces selection problem with the additional nodes is shown in Figure~\ref{fig:kAPRSAGraph}.


Now, the transformed problem can be solved directly using the Hungarian algorithm. The Hungarian algorithm can find the minimum (or maximum) total edge weight in a two-part graph related to a standard allocation problem. For example, if the implemented algorithm is able to find the minimum total weights, but we are looking to find the maximum total weight of the edges, it is enough to negate the weight of years and look for the minimum total edge weight in the new graph.



\subsection{\textbf{Step 2: Obtaining Final Dynamic Optimal Relay Selection and RF interfaces Setting from Solved Assignment Problem}}\label{subsec:solvingJRRSS}


After solving the transformed standard assignment problem, the answer to the main $k$-AP can be achieved according to the obtained solution. In other words, the answer to the main $k$-AP and the new transformed problem can be considered corresponding (the correctness of this correspondence is proved in Appendix \ref{subsec:ProofDORSA}).


After finding the final match between $N_s + N_rN_t$ nodes in both parts of the graph with the maximum total edge weight, the maximum $Q_{BS}$ source will be connected to the previous nodes on the right side of the graph and the remaining nodes will be connected to the new nodes. Therefore, by considering the edges between the sources with the previous nodes on the right side of the graph, relay-RF interface pair or base station, the matching answer of the main problem is obtained. Details on determining the answer will be described in subsection \ref{subsec:DORSA}.

In the next subsection, a pseudo code of the dynamic optimal relay selection and RF interfaces setting algorithm.

\subsection{\textbf{Implementation of Centralized DORSA}}\label{subsec:DORSA}

In this section, pseudo-code related to the implementation of DORSA is provided. As mentioned before, this algorithm has three main steps as follows:  

Step 0: Transforming the joint dynamic relay-RF interfaces selection problem to a $k$-AP
Step 1: Transforming the obtained $k$-AP to a standard assignment problem and solving it
Step 2: Obtaining the final joint dynamic relay-RF interfaces selection solution (from solved standard assignment problem)

To implement steps 0 and 1, we use a two-dimensional matrix to model the bipartite weighted graph related to the joint dynamic relay-RF interfaces selection problem. The rows of the matrix are considered equivalent to the nodes of the left side of the graph (\ref{fig:kAPRSAGraph}) and the columns of the matrix are considered to be equivalent to the nodes of the right side of the graph. The values inside the matrix cells are also equal to the weight of the edge between the corresponding vertices in the row and column of the matrix.

As mentioned earlier, the weight of the edges between:
\begin{itemize}
\item[-] the vertices corresponding to sources and pairs of relays-RF interfaces or the Base station is equal to the data rate between  them,
\item[-] the new vertices and the previous vertices (sources and pairs of relays-RF interfaces or the Base station) of the graph is equal to $A_{value}$,
\item[-] the new nodes on both sides of the graph together is set to zero.
\end{itemize}


Now, the obtained matrix from the new standard assignment problem can be an input of the Hungarian algorithm. It should be noted that where we are looking to find the maximum total network data rate but the used Hungarian algorithm is able to find the minimum value, multiply all the elements of the matrix by $-1$ and then as input to the Hungarian to be given.


In step 2, the output of the Hungarian algorithm is examined and the outputs of the main graph vertices without adding new vertices will be the output of our main $k$-AP. If the content of the answer is less than $N_rN_t$, the desired source is matched to a relay-RF interface pair and the dividend of the matched column number with (the corresponding cell content of) each source on $N_t$ represents the desired relay number and the remainder is equal to the desired RF interface number. Else if $ N_rN_t \leq $ the content of the answer $< N_rN_t + Q_{BS}$, the desired source is matched directly to the base station.




\begin{algorithm}[!htb]
\caption{Proposed Centralized Dynamic Optimal Relay Selection and RF interfaces Setting Algorithm (DORSA).}
\label{alg:DORSA}
\begin{algorithmic}[1]
\algsetup{linenosize=\small}
\scriptsize
  
%

%

  \TITLE{ \textbf{Step 0:Transform the dynamic relay-RF interfaces selection problem to a $k$-AP}} 

	\STATE Construct the first part of the input capacity matrix of the standard assignment problem, $M_{i, j}$, according to the following rules:
	
	 \begin{itemize}	 
		\item[-] $M_{i, j} = min (C_{s,(r,RF_t)}, C_{(r,RF_t), BS}) \quad$	for $(0 \leq i < N_s)$ and $(0 \leq j < N_rN_t)$,  
		\item[-] $M_{i, j} = C_{s,BS}\quad$	for $(0 \leq i < N_s)$ and $(N_rN_t \leq j < N_rN_t+ Q_{BS})$,
	\end{itemize}

  \TITLE{ \textbf{Step 1:Transform the $k$-AP to a standard assignment problem and solve it}} 
    
  \STATE $A_{value} = (max( M_{i, j}) + 1) \times ( N_s+ N_rN_t+Q_{BS}) \quad$ for  $(0 \leq i < N_s)$ and $(0 \leq j < N_rN_t+ Q_{BS})$ \algcost{}{}
  	\STATE Construct the second part of the input capacity matrix of the standard assignment problem, $M_{i, j}$, according to the following rules:
	
	 \begin{itemize}	 
		\item[-] $M_{i, j} = A_{value}\quad$	for $(0 \leq i < N_s)$ and $(N_rN_t+ Q_{BS} \leq j < N_s + N_rN_t)$,
		\item[-] $M_{i, j} =  A_{value}\quad$	for $(N_s \leq i < N_s + N_rN_t)$ and $(0 \leq j < N_rN_t)$.	 
		\item[-] $M_{i, j} = 0\quad$	for $(N_s \leq i < N_s + N_rN_t)$ and $(N_rN_t \leq j < N_s + N_rN_t)$.	 
	\end{itemize}

	\STATE Construct the set of edges $E$ of the bipartite graph by $edge_{i, j}$ = (left node index = $i$, right node index = $j$, $- M_{i, j}$), 
	\STATE $H^o$ vector = $Hungarian_{(FindMin)}$(number of vertices $ = N_s+N_rN_t$, edges = $E$ ),  
			
  \TITLE{ \textbf{Step 2:Obtain the final joint dynamic relay-RF interfaces selection solution}} 
  
  	\STATE Construct the final output assignment vector, $O$, from the output vector of the standard assignment solution, $H^o$, according to the following rules:
	
	\FOR{$k \gets 1$ to $N$}
		\IF{$H^o_k < N_rN_t$} 
			 \item[-] $O_k = H^o_k$: Meaning that the $k$th source is connected to the base station by the relay with index equal to the dividend of $O_k \div N_t$ and the RF interface with index equal to the remainder of $O_k \div N_t$,
		\ELSIF{$ N_rN_t \leq H^o_k < N_rN_t + Q_{BS}$}
			 \item[-] $O_k = N_r$: Meaning that the $k$th source is assigned to the base station directly,
		\ELSIF{$ H^o_k \geq N_rN_t + Q_{BS}$ }
			 \item[-] $O_k = \phi$: Meaning that the $k$th source can not connected to the base station.
		\ENDIF
	\ENDFOR

\end{algorithmic}
\end{algorithm}

The time complexity of Dynamic Optimal Relay Selection and RF interfaces Setting Algorithm (DORSA) is discussed in next subsection.

\subsection{\textbf{Time Complexity of DORSA} }\label{subsec:TimeCompDORSA}

To examine the time complexity of DORSA, we need to examine it. DORSA algorithm is the result of transforming the joint relay selection and dynamic RF interfaces setting problem to a standard assignment problem and solving it by the Hungarian algorithm. Therefore, the resulting time complexity is proportional to the time complexity of the main problem transformation, solving it, and finding its answer. Now we will examine each of the sections separately and briefly.

If we assume that the graph of the main $k$-AP is represented by $G_{k-AP}=(V_{G_{k-AP}},E_{G_{k-AP}})$ and the graph of the transformed standard assignment problem is represented by $G_{sAP}=(V_{G_{sAP}},E_{G_{sAP}})$, we can say:

\begin{itemize}
\item[-] The time complexity of the main problem transformation (step 0 and step 1{lines 2-4}): $O(|E_{G_{k-AP}}|) = O(|V_{G_{sAP}}|^2) = O((N_s+N_rN_t)^2)$,
\item[-] The time complexity of solving transformed standard assignment problem (Hungarian algorithm from step 1{line 5}):$O(|V_{G_{sAP}}||E_{G_{sAP}}|) = O((N_s+N_rN_t)(N_s+N_rN_t)^2) = O((N_s+N_rN_t)^3)$ \cite{DHACMiSt2007, GithFCIH2019}.,
\item[-] The time complexity of finding answer of the main problem: $O(|V_{G_{k-AP}}|)= O(N_s)$.
\end{itemize}

Therefore, the time complexity of DORSA is equal to $O(n^3)$ where $n = N_s+N_rN_t$. Thus, if two of the three parameters are constant, the time complexity is proportional to the third power of the third parameter.





\section{Simulation Results}\label{sec:simulation}
	
We are simulating our scenarios in M2MSim. M2MSim is a simulator written by C++. This version of M2MSim is an extension of the initial code of the relay selection algorithm with static RF interfaces setting \cite{MRSRGhHe2020} for dynamic RF interfaces setting in M2M communications. In this extension, we considered the possibility of using multiple RF interfaces between machines in some algorithms. The considered RF interfaces for M2M include WiFi, Bluetooth, and Z-Wave, but the implemented RF interface for M2B, like the previous version, is LTE. The simulations of our scenarios were executed on a device with a 4-core Intel Xeon CPU (2.49 GHz) and 4 GB of RAM. 



The scenario environment was a square with its size is equal to $500 \times 500 (m^2)$. In this environment, $N$ machine ($N_s$ sources $+$ $N_r$ relays) are placed randomly with a uniform distribution. Each scenario was run $n=200$ times and the average of results were considered. The details of simulation parameters are provided in Table \ref{tbl:SimPar}.

Note: To calculate the SINR in the simulations, the highest level of interference is considered. So in the real world, simulation results can be the least we can expect from the results of calculating the total data rate.




\color{red}


\begin{table}
   \scriptsize
   \renewcommand{\arraystretch}{1.3}
  \begin{center}
    \caption{Simulation Parameters.}
    \label{tbl:SimPar}
    \begin{tabular}{|l|c|} 
    \hline
      \textbf{Parameter} & \textbf{Default Value} \\
      \hline
      WiFi Uplink Central Frequency & $5600 \mathrm{MHz}$ \\
      WiFi Total Bandwidth & $20 \mathrm{MHz}$ \\ 
      WiFi Power Transmission of Machines & $0.1 w$ \\
      WiFi Power Transmission of Base station & $0.1 w$ \\
      WiFi Max Data Rate for each device & $54 \mathrm{Mbps}$ \\
      WiFi Max Distance Range & $120 m$ \\
      Bluetooth Uplink Central Frequency & $2400 \mathrm{MHz}$ \\
      Bluetooth Total Bandwidth & $1 \mathrm{MHz}$ \\
      Bluetooth Power Transmission of Machines & $2.5 mw$ \\
      Bluetooth Power Transmission of Base station & $2.5 mw$ \\
      Bluetooth Max Data Rate for each device & $3 \mathrm{Mbps}$ \\
      Bluetooth Max Distance Range & $10 m$ \\
      Z-Wave Uplink Central Frequency & $908.42 \mathrm{MHz}$ \\
      Z-Wave Total Bandwidth & $200 \mathrm{KHz}$ \\
      Z-Wave Power Transmission of Machines & $1 mw$ \\
      Z-Wave Power Transmission of Base station & $1 mw$ \\
      Z-Wave Max Data Rate for each device & $100 \mathrm{Kbps}$ \\
      Z-Wave Max Distance Range & $30 m$ \\
      LTE Uplink Central Frequency & $1910 \mathrm{MHz}$ \\
      LTE Total Bandwidth & $20 \mathrm{MHz}$\\
      Max Number of LTE Channel ($No^{ch}_{\mathrm{LTE}}$) & $128$ \\
      LTE Power Transmission of Machines & $0.2 w$ \\
      LTE Power Transmission of Base station & $10 w$ \\
      LTE Max Data Rate for each device & $100 \mathrm{Mbps}$ \\
      LTE Max Distance Range & $1 Km$ \\
      Mean of Normal Shadowing on Received Power & $0$ \\
      Std. Dev. of Normal Shadowing on Received Power & $8$ \\ 
      Number of Simulation Runs & $200$ \\
      Number of Machines ($N$) & $(Default:) \quad 120$ \\
      Number of Sources ($N_s$) & $0 \quad to \quad 120$ \\
      Number of Relays ($N_r$) & $N-N_s$ \\
      Length of Test Environment & $500 m$  \\
     Width of Test Environment & $500 m$ \\
      \hline
    \end{tabular}
  \end{center}
\end{table}

\color{black}

In different scenarios, one or more types of algorithms are compared in different situations. Direct Transmission with Optimal next hop Selection Algorithm (DiTOSA) is an optimal selection algorithm that matches the sources to the base station directly via LTE for M2B communications. This matching is such that in the end, according to the amount of source requests, the maximum total data rate in the network is obtained.


The method for calculating the time complexity of the DiTOSA is similar to DORSA‌ except that the number of vertices of the original and transformed graphs is different. In other words, in DiTOSA the number of relays ($N_r$) and RF interfaces ($N_t$) is zero. Thus the time complexity of DiTOSA is equal to $O({n_1}^3)=O({N_s}^3)$. So, the time complexity of DiTOSA depends only on the number of sources.


Static Optimal Relay Selection Algorithm (SORSA) is a relay selection algorithm with static RF interfaces setting with one RF interface for M2M communications and LTE for M2B communications \cite{MRSRGhHe2020}.

The time complexity of SORSA is equal to $O({n_2}^3) = O((N_s+N_r)^3)$, as discussed in related paper. Therefore, the time complexity of SORSA depends on the number of sources and relays.

\begin{itemize}
\item[-] SORSA\_W: SORAS with Wifi for M2M communications,
\item[-] SORSA\_B: SORAS with Bluetooth for M2M communications, and
\item[-] SORSA\_Z: SORAS with Z-Wave for M2M communications.

\end{itemize}

In the simulations, DORSA was also tested under different conditions, with multiple RF interfaces for M2M communications and LTE for M2B communications.
\begin{itemize}
\item[-] DORSA\_W-B: DORAS with Wifi and Bluetooth for M2M communications,
\item[-] DORSA\_W-Z: DORAS with Wifi and Z-Wave for M2M communications,
\item[-] DORSA\_B-Z: DORAS with Bluetooth and Z-Wave for M2M communications,
\item[-] DORSA\_W-B-Z: DORAS with Wifi, Bluetooth, and Z-Wave for M2M communications.
\end{itemize}


The time complexity of DORSA is equal to $O({n_3}^3) = O((N_s+N_rN_t)^3)$, as discussed in subsection \ref{subsec:TimeCompDORSA}. Thus, DORSA time complexity depends on the number of sources, relays, and M2M RF interfaces.

Therefore, if there is no relay, the execution time of DiTOSA, SORSA, and DORSA are similar to each other, otherwise, if there are relays with one interface, regardless of the type of interface, the execution time of  SORSA‌ and DORSA‌ are similar to each other.
Otherwise, if the number of M2M RF interfaces is more than one, the execution time of DORSA‌ will increase in proportion to the power of three of the number of M2M RF interfaces.


To compare the performance of different algorithms in different conditions, the following parameters were examined:

\begin{itemize}
\item[-] \textbf{Average data rate of connections between Sources and the Base station}:  Given that in the model system, uplink connections are examined, the average data rate possible for data exchange by sources in the matching conditions performed by each of the algorithms is considered as a parameter. In the following, we will briefly call this parameter "data rate".


\item[-] \textbf{Average Number of Unmatched Sources}: The average number of sources by the selection algorithm did not match. This parameter is briefly called "unmatched source number".


\item[-] \textbf{Actual Execution Time of Proposed Algorithms}: The average execution time of each algorithm in different conditions is checked by this parameter. The upper bound of execution time will be at most proportional to the order of time complexity, that discussed earlier. We will briefly call this parameter "actual execution time".

\end{itemize}

In the following, the conditions related to the different scenarios examined are stated.

\begin{itemize}
\item[-] \textbf{Scenario 1}: This scenario was investigated in order to simulate an almost real uplink cell in which some machines have data (sources) and some do not have data and can act as relays. Therefore, in this scenario, the number of \textbf{machines} is assumed to be \textbf{constant} and the number of \textbf{relays} \textbf{decreases} as the number of \textbf{sources} \textbf{increases}.



\item[-] \textbf{Scenario 2}: In this scenario, to check the change in the number of sources in the parameters under consideration, the number of \textbf{relays} is assumed to be \textbf{constant} and the number of \textbf{sources} in the desired range is \textbf{changed}.



\item[-] \textbf{Scenario 3}: In this scenario, to check the change in the number of relays in the parameters under consideration, the number of \textbf{sources} is assumed to be \textbf{constant} and the number of \textbf{relays} in the desired range is \textbf{changed}.



\item[-] \textbf{Scenario 4}: In this scenario, to check the change in the volume of source requests in the parameters under consideration, the number of \textbf{sources} and \textbf{relays} are assumed to be \textbf{constant} and \textbf{the requested bandwidth by the sources} in the desired range is \textbf{changed}.



\end{itemize}

In the following, we examine each of the scenarios separately.



%
%
%
%
%
%
%
%
%

	\subsection{\textbf{Scenario 1}}\label{sc:S1}

Scenario 1 examines changing the number of sources and relays while the number of machines is constant in DiTOSA, SORSA\_W, SORSA\_B, SORSA\_Z,	DORSA\_W-B, DORSA\_W-Z, DORSA\_B-Z, and DORSA\_W-B-Z. The simulation parameters of scenario 1 are provided in Table \ref{tbl:s1_SimPar}.

\begin{table}[H]
   \scriptsize
   \renewcommand{\arraystretch}{1.3}
  \begin{center}
    \caption{Simulation Parameters of Scenario 1.}
    \label{tbl:s1_SimPar}
    \begin{tabular}{|l|c|c|} 
    \hline
      \textbf{Parameter} & \textbf{Value} & \textbf{Constant/Variable}\\
      \hline
      \hline
      $N$ & $240$ & Constant\\
      $N_s$ & $0 .. 120$ & Variable\\
      $N_r$ & $N-N_s$ & Variable\\
      $RBW_s$ & $200 KHz$ & Constant\\
      \hline
    \end{tabular}
  \end{center}
\end{table}
where the requested bandwidth by the sources is called by $RBW_s$.

 \begin{enumerate}

	\item \textbf{Data Rate}
			
After reviewing the results of the data rate chart (Figure~\ref{fig:s1Cap}), a few points can be mentioned, which we will briefly describe below. As can be seen from the results, there are two important points:
	
\begin{enumerate}
\item Using relays with more M2M RF interfaces can provide a higher average data rate than without relays or with fewer RF interfaces, assuming they have the same intermediates.
\item The use of more RF interfaces alone does not necessarily improve the average data rate of the network. In addition to the number, the type of RF interfaces and data rates that can provide the average bandwidth requested by network resources are also important.
\end{enumerate}

%
%
	
Let us examine in more detail:	
It appears that all \\ DORSA\_W-B, DORSA\_W-Z, DORSA\_B-Z, \\ and DORSA\_W-B-Z algorithms performed better than DiTOSA. Furthermore, it can be seen that the presence of the WiFi interface in SORSA\_W has resulted in a higher data rate than the DORSA\_B-Z data rate. Although a review of the results of DORSA\_W-B-Z in this scenario can show better results than all other algorithms, algorithms such as DORSA\_W-B, DORSA\_W-Z, and even SORSA\_W were able to provide a slightly different data rate than DORSA\_W-B-Z, which had the best average. But DORSA\_B-Z still outperformed SORSA\_B and SORSA\_Z on average. Finally, DORSA‌\_W-B-Z is on average 10\% better than other algorithms, up to 0.6\% better than SORSA\_W, and up to 74\% better than DiTOSA.

Therefore, in addition to increasing the number of communication intermediaries that can be effective in the average network data rate, their type can also affect the performance of the algorithm.


95\% confidence interval of results are CI[$( 1.05 \times 10^6) \pm 0.17\%$], CI[$( 1.26 \times 10^6) \pm 0.10\%$], CI[$( 1.09 \times 10^6) \pm 0.15\%$], CI[$( 1.06 \times 10^6) \pm 0.16\%$], CI[$( 1.26 \times 10^6) \pm 0.10\%$], CI[$( 1.26 \times 10^6) \pm 0.10\%$], CI[$( 1.10 \times 10^6) \pm 0.15\%$], and CI[$( 1.26 \times 10^6) \pm 0.10\%$] for DiTOSA, \\ SORSA\_W, SORSA\_B, SORSA\_Z, DORSA\_W-B, \\ DORSA\_W-Z, DORSA\_B-Z, and DORSA\_W-B-Z, respectively.

			\begin{figure}
				\centering
				\includegraphics[scale=0.5]{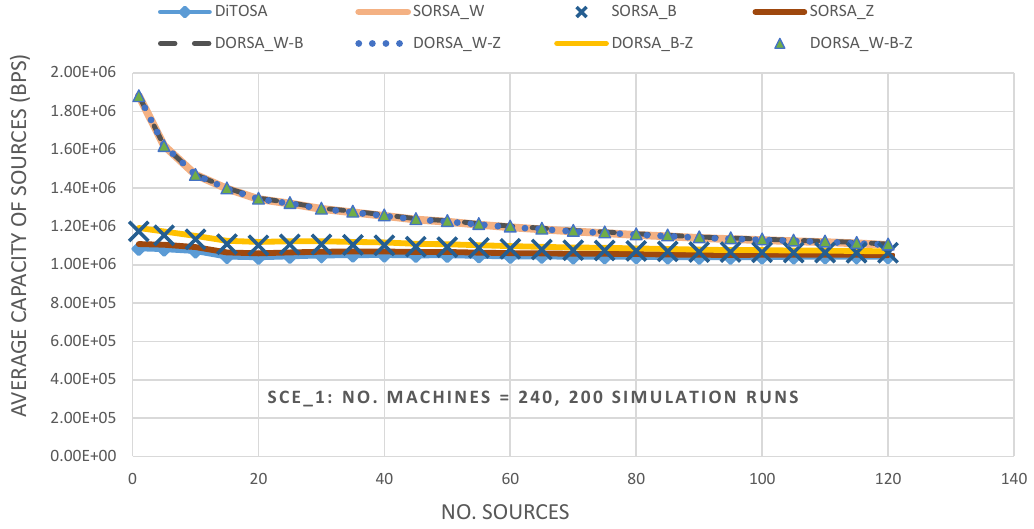}
				\caption[The average data rate of sources for all algorithms (bps) vs. the number of sources in Scenario 1 with 240 machines and 200 simulation runs.]{The average data rate of sources for all algorithms (bps) vs. the number of sources in Scenario 1 with 240 machines and 200 simulation runs.}
				\label{fig:s1Cap}
			\end{figure}

	\item \textbf{Unmatched Source Number}

The maximum number of sources connected to the base station, directly or by relay, is equal to the product of the total LTE bandwidth allocated to the base station based on the amount of source requests in that network cell.


Therefore, the number of sources that can not be connected to the base station is at least equal to the total number of network cell sources minus the maximum number of sources that can be connected due to empty channel communication capacity with the base station.


Max number of connected sources, directly or by 2 hops is equal to: \\
 $ \frac{BW^{(LTE)}_{Total}}{BW_{ReqSources}} =\frac{20 MHz}{200 KHz} = 100$
where $BW^{(LTE)}_{Total}$ is the total BW of LTE and $BW_{ReqSources}$ is the requested BW of sources.
So, the base station can accept up to 100 machines, sources, or relays. As shown in Figure~\ref{fig:s1Unmatched}, as long as the number of sources is less than or equal to 100, the number of untapped sources is zero. Then by adding as many sources as possible from 100 sources, the number of sources that could not connect to the base station, directly or by relays, is equal to the difference between the total number of sources and 100 matched sources.


		

%
%
%


			\begin{figure}
			\centering
			\includegraphics[scale=0.5]{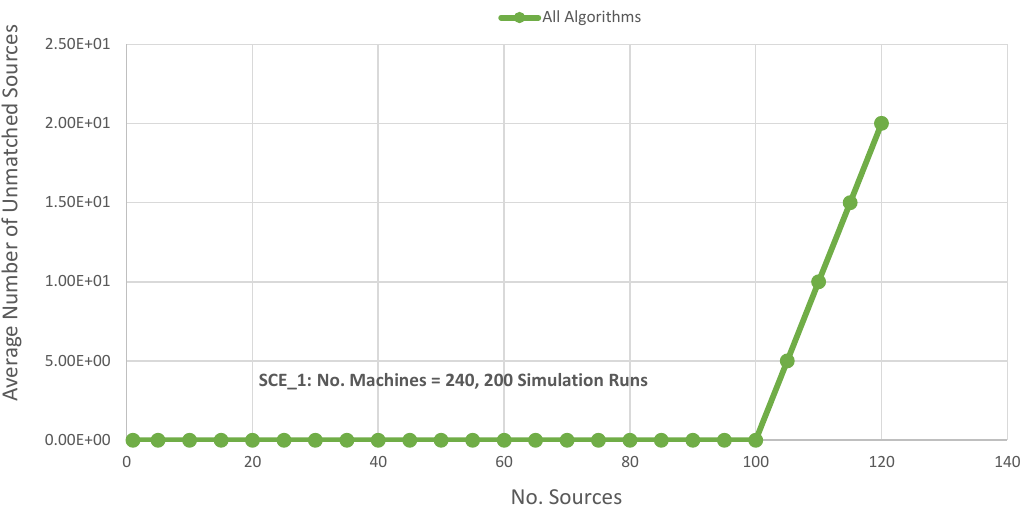}
			\caption[The average number of unmatched sources for all algorithms vs. the number of sources in Scenario 1 with 240 machines and 200 simulation runs.]{The average number of unmatched sources for all algorithms vs. the number of sources in Scenario 1 with 240 machines and 200 simulation runs.}
			\label{fig:s1Unmatched}
			\end{figure} 


		\item \textbf{Actual Execution Time}
		
In this subsection, the execution times of different algorithms are examined together. In general, the execution times are relatively close to each other, but especially at the beginning of the chart, when the number of relays is more and at the same time the number of interfaces is higher, DORSA\_W-B-Z had more execution time than the rest.

As the number of sources increases and as a result, the number of relays decreases, as mentioned before, the execution time diagram of all algorithms gets closer to each other. The next scenarios focus on the execution time of DORSA.


On the other hand, it can be seen that in places with more than 100 sources, due to the fact that several sources do not match, it takes longer to find sources whose matching maximizes the average data rate.

As shown in Figure~\ref{fig:s1time}, the maximum time complexity is equal to 3 (as previously discussed) and the execution of the algorithms took a maximum of 544 (ms) and a minimum of 162 (ms).


		

			
		\begin{figure}
			\centering
			\includegraphics[scale=0.47]{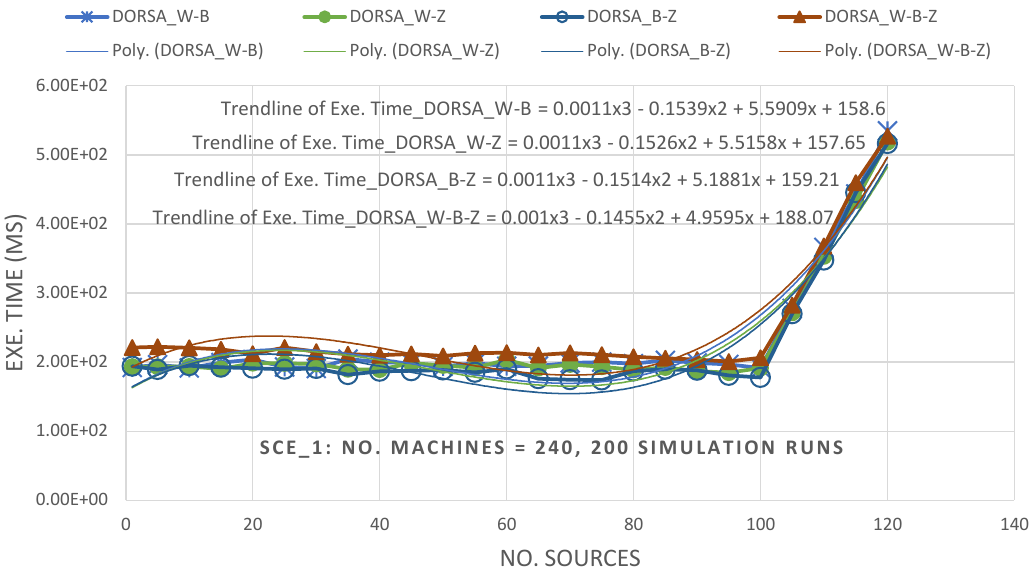}
			\caption[The average actual execution time (ms) and its trendline for DORSA\_W-B, DORSA\_W-Z, DORSA\_B-Z, and DORSA\_W-B-Z vs. the number of sources in Scenario 1 with 240 machines and 200 simulation runs.]{The average actual execution time (ms) and its trendline for DORSA\_W-B, DORSA\_W-Z, DORSA\_B-Z, and DORSA\_W-B-Z vs. the number of sources in Scenario 1 with 240 machines and 200 simulation runs.}
			\label{fig:s1time}
		\end{figure} 			
				
		\end{enumerate}
		

			\subsection{\textbf{Scenario 2}}\label{sc:S2}

In this scenario, the effect of changing the number of sources in the algorithms is investigated in conditions where parameters such as the number of relays and the amount of bandwidth requested by the sources are constant. The simulation parameters of scenario 2 are provided in Table \ref{tbl:s2_SimPar}.


The difference between Scenario 2 and Scenario 1 is that unlike Scenario 1, where the number of machines was constant and equal to 240, the number of relays changed from 240 to 120 by changing the number of sources from zero to 120, in this scenario, the number of relays will always be constant and equal to 120.


\begin{table}[H]
   \scriptsize
   \renewcommand{\arraystretch}{1.3}
  \begin{center}
    \caption{Simulation Parameters of Scenario 2.}
    \label{tbl:s2_SimPar}
    \begin{tabular}{|l|c|c|} 
    \hline
      \textbf{Parameter} & \textbf{Value} & \textbf{Constant/Variable}\\
      \hline
      \hline
      $N$ & $N_s+N_r$ & Variable\\
      $N_s$ & $1 .. 120$ & Variable\\
      $N_r$ & $120$ & Constant\\
      $RBW_s$ & $200 KHz$ & Constant\\
      \hline
    \end{tabular}
  \end{center}
\end{table}
where the requested bandwidth by the sources is called by $RBW_s$.

		\begin{enumerate}

			\item \textbf{Data Rate}

As shown in Figure~\ref{fig:s2Cap}, the diagram overview of the average data rate of all algorithms in this scenario is almost similar to Scenario 1. Only concerning the number of relays less or equal in the diagram of Scenario 2 compared to Scenario 1, we see the average data rate less than equal to Scenario 1.

In this scenario, similar to the previous scenario, the \\ DORSA\_W-B-Z algorithm is better than the other algorithms, and DORSA\_W-B, DORSA\_W-Z, and SORSA\_W are slightly behind DORSA\_W-B-Z due to the use of M2M RF interfaces and the use of WiFi bandwidth. In addition, DiTOSA, which does not use relays and  M2M RF interfaces, achieved a lower average data rate, albeit with a slight difference from other algorithms.

In this scenario, similar to Scenario 1, the data rate of DORSA\_W-B-Z algorithm is on average 8\% better than other algorithms, up to about 0.5\% better than SORSA\_W, and up to 74\% better than DiTOSA algorithm.




95\% confidence interval of results are CI[$(1.04 \times 10^6) \pm 0.16\%$], CI[$(1.20 \times 10^6) \pm 0.10\%$], CI[$(1.06 \times 10^6) \pm 0.15\%$], CI[$(1.05 \times 10^6) \pm 0.15\%$], CI[$( 1.21 \times 10^6) \pm 0.10\%$], CI[$( 1.21 \times 10^6) \pm 0.10\%$], CI[$( 1.08 \times 10^6) \pm 0.15\%$], and CI[$( 1.21 \times 10^6) \pm 0.10\%$] for DiTOSA, SORSA\_W, SORSA\_B, SORSA\_Z, DORSA\_W-B, DORSA\_W-Z, DORSA\_B-Z, and DORSA\_W-B-Z, respectively.





			\begin{figure}
				\centering
				\includegraphics[scale=0.5]{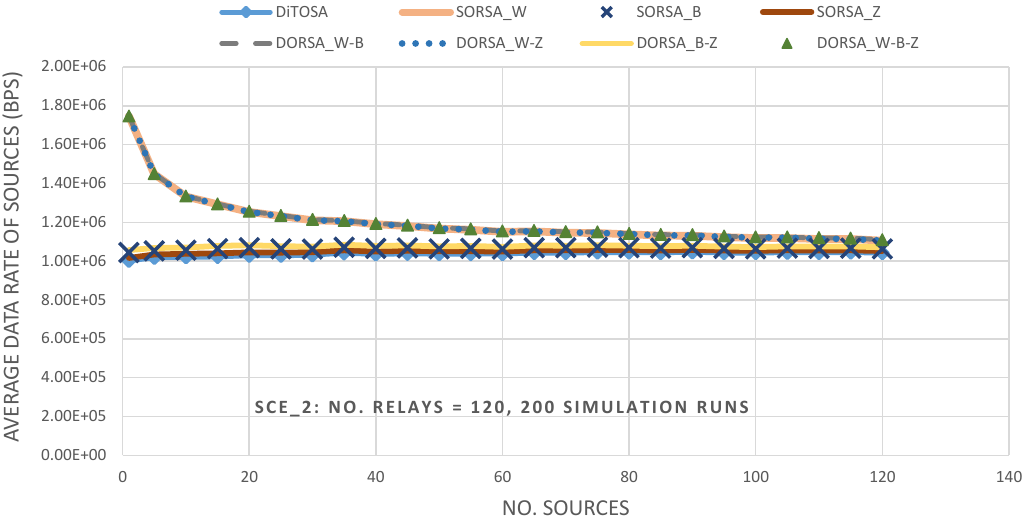}
				\caption[The average data rate of sources for all algorithms (bps) vs. the number of sources in Scenario 2 with 120 relays and 200 simulation runs.]{The average data rate of sources for all algorithms (bps) vs. the number of sources in Scenario 2 with 120 relays and 200 simulation runs.}
				\label{fig:s2Cap}
				\end{figure}




		\item \textbf{Unmatched Source Number}

Due to the similarity of the requested BW of the sources and the number of sources with scenario 1, it was also observed in this section that as long as the number of sources is less than or equal to 100, all sources are matched and the number of unmatched sources is equal to zero. Then, as the number of network cell sources differs from the maximum number of sources to which the base station can be connected ( $=$ 100 sources), the number of unmatched sources increases to 20 sources.


(Note: Due to the same number of unmatched sources in all cases with Scenario 1, the relevant chart was not displayed and was enough to explain)

			

\item \textbf{Actual Execution Time}

In this subsection, we focused specifically on the execution time of DORSA algorithms. Figure~\ref{fig:s2time} shows the actual execution time of DORSA\_W-B, DORSA\_W-Z, DORSA\_B-Z, and DORSA\_W-B-Z. This time is calculated as a maximum of 627 (ms) and a minimum of 66 (ms).

c which confirms the correctness of the order $O(n^3)$ ‌ previously calculated in this scenario.





				\begin{figure}
			\centering
			\includegraphics[scale=0.47]{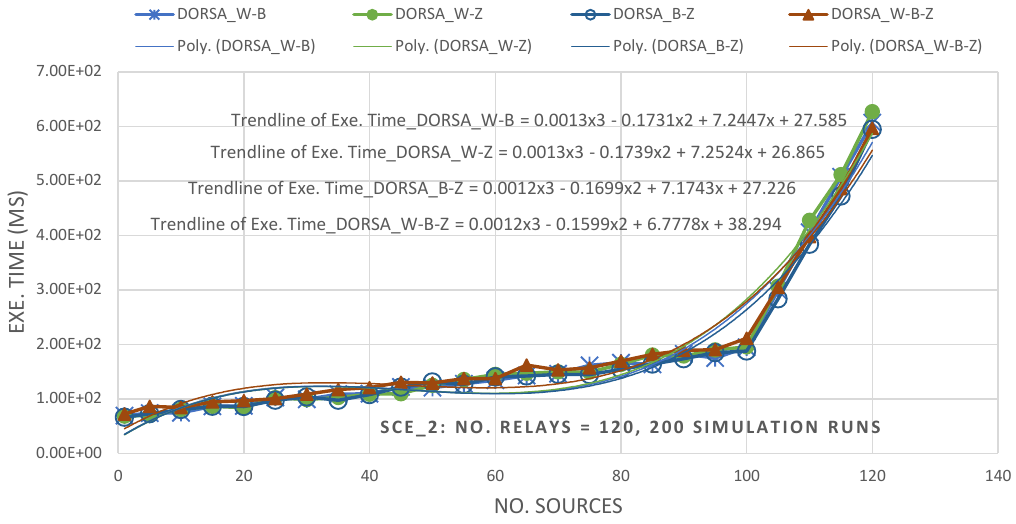}
			\caption[The average actual execution time (ms) and its trendline for DORSA\_W-B, DORSA\_W-Z, DORSA\_B-Z, and DORSA\_W-B-Z vs. the number of sources in Scenario 2 with 120 relays and 200 simulation runs.]{The average actual execution time  (ms) and its trendline for DORSA\_W-B, DORSA\_W-Z, DORSA\_B-Z, and DORSA\_W-B-Z vs. the number of sources in Scenario 2 with 120 relays and 200 simulation runs.}
			\label{fig:s2time}
			\end{figure}

		\end{enumerate}

	\subsection{\textbf{Scenario 3}}\label{sc:S3}		
The simulations in Scenario 3 focus specifically on changing the number of relays. In this scenario, by keeping the number of sources constant, the effect of changing the number of relays is investigated. The simulation parameters in this scenario are shown in Table \ref{tbl:s3_SimPar}.		
\begin{table}[htb]
   \scriptsize
   \renewcommand{\arraystretch}{1.3}
  \begin{center}
    \caption{Simulation Parameters of Scenario 3.}
    \label{tbl:s3_SimPar}
    \begin{tabular}{|l|c|c|} 
    \hline
      \textbf{Parameter} & \textbf{Value} & \textbf{Constant/Variable}\\
      \hline
      \hline
      $N$ & $N_s+N_r$ & Variable\\
      $N_s$ & $120$ & Constant\\
      $N_r$ & $1 .. 120$ & Variable\\
      $RBW_s$ & $200 KHz$ & Constant\\
      \hline
    \end{tabular}
  \end{center}
\end{table}
where the requested bandwidth by the sources is called by $RBW_s$.
				\begin{enumerate}
					\item \textbf{Data Rate}

Figure~\ref{fig:s3Cap} shows the average data of different algorithms in a situation where the number of sources is constant and equal to 120 and the number of relays varies between 0 and 120. As shown in the figure, as in the previous scenarios, DiTOSA averaged the lower data rate and DORSA\_W-B-Z reached the highest average data rate. In other words, it can be said that the following trend is established in different parts of the diagram between different algorithms:
$DiTOSA  <  SORSA\_Z < SORSA\_B <  DORSA\_B-Z     <     SORSA\_W  \leq  DORSA\_W-Z \leq DORSA\_W-B \leq DORSA\_W-B-Z$
In this scenario, DORSA\_W-B-Z performed on average about 2\% better than other algorithms, ,up to about 0.5\% better than SORSA\_W, and in most cases about 7\% better than DiTOSA. The standard deviation of algorithms for all algorithms are equal to 0.08 in all cases, so the 95\% confidence interval of results are equal to CI[$( 1.04 \times 10^6) \pm 0.08\%$], CI[$( 1.08\times 10^6) \pm 0.08\%$], CI[$( 1.05 \times 10^6) \pm 0.08\%$], CI[$( 1.05 \times 10^6) \pm 0.08\%$], CI[$( 1.08 \times 10^6) \pm 0.08\%$], CI[$( 1.08 \times 10^6) \pm 0.08\%$], CI[$( 1.06 \times 10^6) \pm 0.08\%$], and CI[$( 1.08 \times 10^6) \pm 0.08\%$] for DiTOSA, SORSA\_W, SORSA\_B, SORSA\_Z, DORSA\_W-B, DORSA\_W-Z, DORSA\_B-Z, and DORSA\_W-B-Z, respectively.
			\begin{figure}
				\centering
				\includegraphics[scale=0.5]{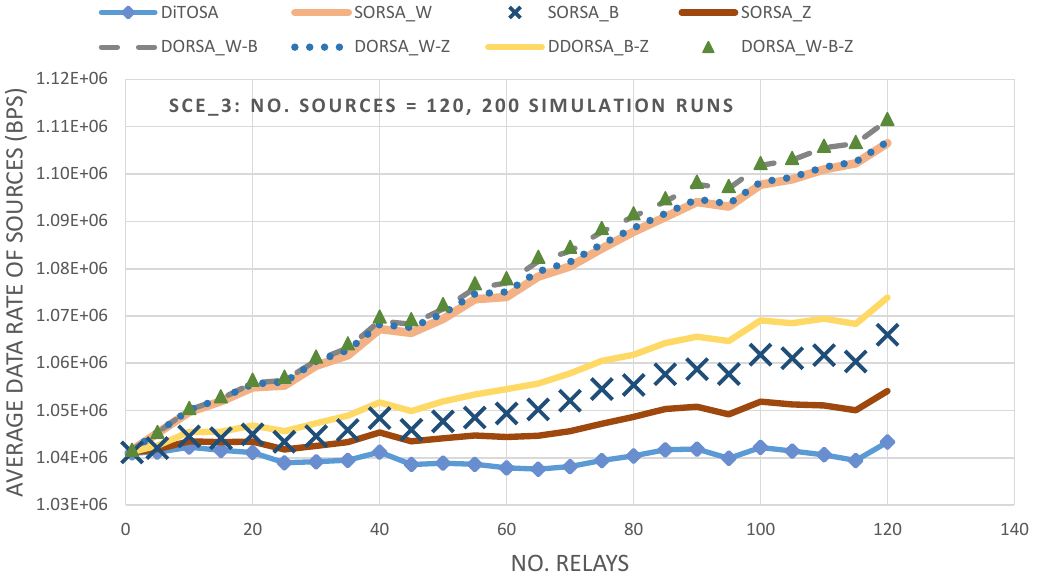}
				\caption[The average data rate of sources for all algorithms (bps) vs. the number of relays in Scenario 3 with 120 sources and 200 simulation runs.]{The average data rate of sources for all algorithms (bps) vs. the number of relays in Scenario 3 with 120 sources and 200 simulation runs.}
				\label{fig:s3Cap}
				\end{figure} 		
	\item \textbf{Unmatched Source Number}
	
Given that in this scenario the number of sources in all conditions is equal to 120, but as mentioned before, the maximum number of sources to which the base station can be connected is equal to 100, in all conditions 20 sources do not match.
	
	(Note: Due to the same number of unmatched sources in all cases, the relevant chart was not displayed and was enough to explain)


\item \textbf{Actual Execution Time}

In this scenario, the execution time of DORSA algorithms can be seen in Figure~\ref{fig:s3time}. As can be seen, in the case of fixed sources, the actual execution time is approximately linear and still does not violate the upper bound of time complexity when previously calculated. In this case, the maximum execution time is 473 (ms) and the minimum execution time is 330 (ms).
				\begin{figure}
			\centering
			\includegraphics[scale=0.47]{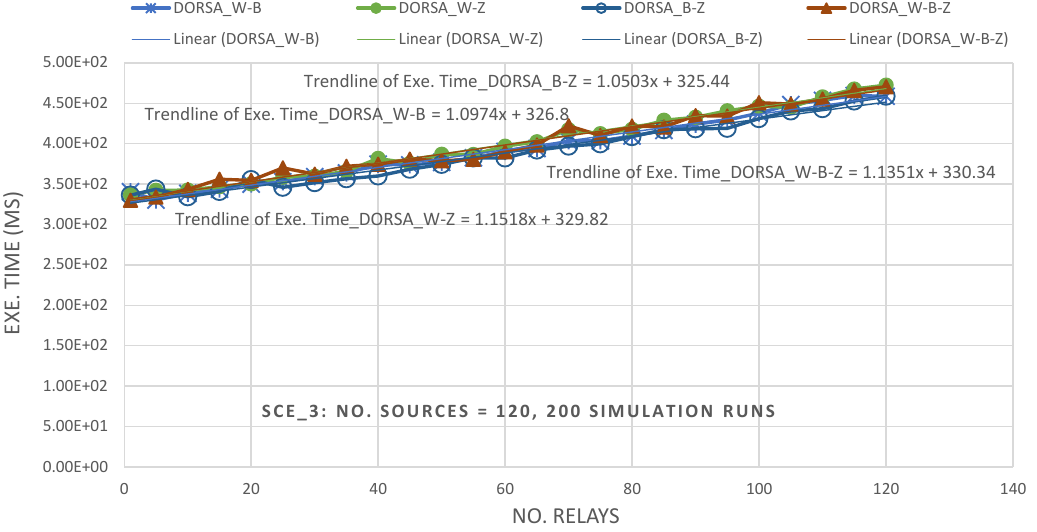}
			\caption[The average actual execution time (ms) and its trendline for DORSA\_W-B, DORSA\_W-Z, DORSA\_B-Z, and DORSA\_W-B-Z vs. the number of relays in Scenario 3 with 120 sources and 200 simulation runs.]{The average actual execution time  (ms) and its trendline for DORSA\_W-B, DORSA\_W-Z, DORSA\_B-Z, and DORSA\_W-B-Z vs. the number of relays in Scenario 3 with 120 sources and 200 simulation runs.}
			\label{fig:s3time}
			\end{figure} 			
				\end{enumerate}

	\subsection{\textbf{Scenario 4}}\label{sc:S4}

This scenario examines the effect of the requested bandwidth of the sources if the number of machines is constant. Table \ref{tbl:s4_SimPar} states the simulation parameters of Scenario 4.

\begin{table}[H]
   \scriptsize
   \renewcommand{\arraystretch}{1.2}
  \begin{center}
    \caption{Simulation Parameters of Scenario 4.}
    \label{tbl:s4_SimPar}
    \begin{tabular}{|l|c|c|} 
    \hline
      \textbf{Parameter} & \textbf{Value} & \textbf{Constant/Variable}\\
      \hline
      \hline
       $N$ & $N_s+N_r$ & Constant\\
      $N_s$ & $120$ & Constant\\
      $N_r$ & $120$ & Constant\\
      $RBW_s$  & $ \lbrace 20, 100, 200, 400, 600, 800, 1000, 2000,$ & Variable\\
      & $ 6000, 10000, 15000 , 20000 \rbrace (KHz)$ & \\
      \hline
    \end{tabular}
  \end{center}
\end{table}	
where the requested bandwidth by the sources is called by $RBW_s$.


				\begin{enumerate}
		
						\item \textbf{Data Rate}

As shown in Figure~\ref{fig:s4Cap}, the average data rate of the algorithms does not change much with the change in the requested bandwidth of the sources and the results of the algorithms are slightly different from each other. Although in this case, the DORSA\_W-B-Z algorithm has an average result of 0.8\% better than other algorithms, up to 6\% better than SORSA\_W, and in the case of maximum difference, it performed 8\% better than DiTOSA.

Now we describe the chart trend for the different requested bandwidth of sources:

\begin{enumerate}
\item[-] Point 1: In this point of diagrams, each source request 20 KHz, so all three M2M interfaces (WiFi, Bluetooth, Z-Wave) and one M2B interface (LTE) are usable. As a result, the total requested bandwidth by 120 sources is less than the total available bandwidth for the base station, and all of them can be connected to the base station directly or by relay.    


\item[-] Point 2: In this point of diagrams, each source request 100 KHz, so all three M2M interfaces (WiFi, Bluetooth, Z-Wave) and one M2B interface (LTE) are usable. As a result, at this point as well as at the previous point with an uptrend, the total requested bandwidth by 120 sources is less than the total available bandwidth for the base station, and all of them can be connected to the base station directly or by relay. 

\item[-] Point 3: In this point of diagrams, each source request 200 KHz, so all three M2M interfaces (WiFi, Bluetooth, Z-Wave) and one M2B interface (LTE) are usable. Therefore, as the chart continues to rise, the total requested bandwidth by 120 sources is greater than the total bandwidth available for the base station, and only a maximum of 100 sources can be connected to the base station directly or by relay. 


\item[-] Point 4: In this point of diagrams, each source request 400 KHz, therefore, two types of M2M interface (WiFi, Bluetooth) and one M2B interface (LTE) can be used. Therefore, as the chart continues to rise as before, the total bandwidth requested by all sources is greater than the total bandwidth available for the base station, and a maximum of 50 sources can be connected to the base station directly or by relay. 


\item[-] Point 5: In this point of diagrams, each source request 600 KHz, so, two types of M2M interface (WiFi, Bluetooth) and one M2B interface (LTE) can be used. As a result, as before, as the uptrend continues, the total bandwidth requested by all sources is greater than the total available bandwidth for the base station, and a maximum of 30 sources can be connected to the base station directly or by relay.


\item[-] Point 6: In this point of diagrams, each source request 800 KHz, so, two types of M2M interface (WiFi, Bluetooth) and one M2B interface (LTE) can be used. Therefore, in this point as in the previous points and as the uptrend continues, not all sources can be connected to the base station directly or by relay, and a maximum of 25 sources can be connected to the base station.


\item[-] Point 7: In this point of diagrams, each source request 1 MHz, therefore, two types of M2M interface (WiFi, Bluetooth) and one M2B interface (LTE) can be used. Therefore, as before, as the uptrend continues, not all sources can connect to the base station, and a maximum of 20 sources can be connected to the base station.


\item[-] Point 8: In this point of diagrams, each source request 2 MHz, so only one type of M2M interface (WiFi) and one M2B interface (LTE) can be used. As a result, as in the previous points and as the uptrend continues, not all sources can be connected to the base station, and a maximum of 10 sources can be connected to the base station.


\item[-]  Point 9: In this point of diagrams, each source request 2 MHz, so only one type of M2M interface (WiFi) and one M2B interface (LTE) can be used. Therefore, as in the previous points and with the continuation of the upward trend, although with a small difference, not all sources can be connected to the base station and a maximum of 3 sources can be connected to the base station.


\item[-]  Point 10: In this point of diagrams, each source request 10 MHz, so only one type of M2M interface (WiFi) and one M2B interface (LTE) can be used. Therefore, at this point, not all sources can be connected to the base station, and a maximum of 2 sources with the mentioned requested bandwidth and a maximum data rate of 100 Mbps can be connected to the base station. Since the total data rate of both sources in this case divided by the total number of sources is less than in the previous case, the chart trend has decreased.



\item[-]  Point 11: In this point of diagrams, each source request 15 MHz, so only one type of M2M interface (WiFi) and one M2B interface (LTE) can be used. At this point, not all sources can be connected to the base station, and a maximum of 1 source with maximum data rate of 100 (Mbps) can be connected to the base station. As a result, the process of reaching this point is lower than the previous point. 


\item[-]  Point 12: In this point of diagrams, each source request 20 MHz, so only one type of M2M interface (WiFi) and one M2B interface (LTE) can be used. Therefore, as in the previous point, not all sources can be connected to the base station, and a maximum of 1 source with maximum data rate of 100 (Mbps) can be connected to the base station.


\end{enumerate}

Note: If the chart continued, none of the sources could be serviced because the requested bandwidth was not supported by LTE.


The 95\% confidence interval of results for all selection algorithms is equal to  CI[$(1.4 \times 10^6) \pm 0.05\%$] in all cases.

.


%

			\begin{figure}
				\centering
				\includegraphics[scale=0.5]{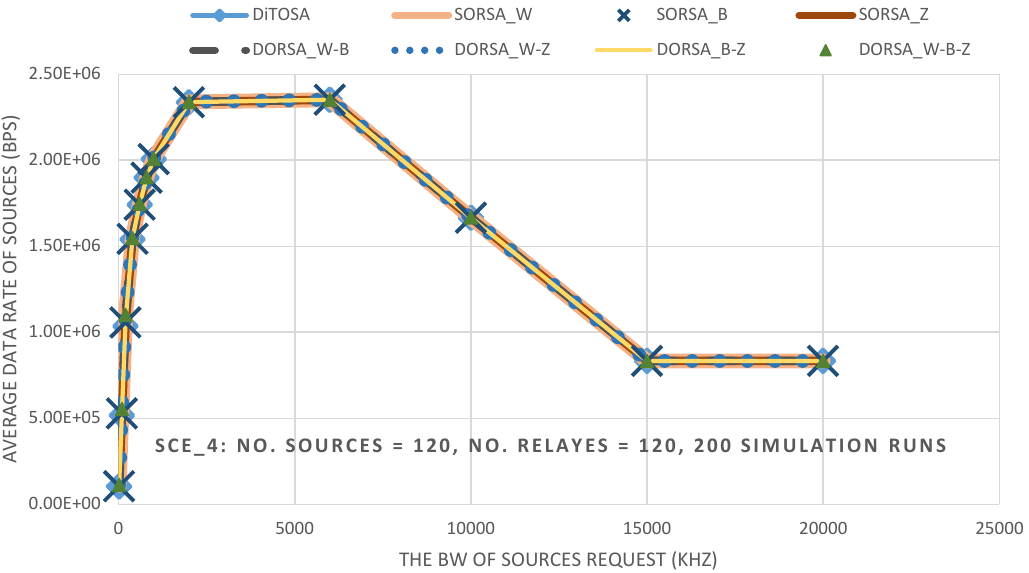}
				\caption[The average data rate of sourcesfor all algorithms (bps) vs. the BW of sources request (kHz) in Scenario 4 with 120 sources and relays and 200 simulation runs.]{The average data rate of sources for all algorithms (bps) vs. the BW of sources request (kHz) in Scenario 4 with 120 sources and relays and 200 simulation runs.}
				\label{fig:s4Cap}
			\end{figure}
			\item \textbf{Unmatched Source Number}

As described in the previous section, all sources can be connected at the first two points of the graph, so the number of unmatched sources is initially zero. Then, as the requested BW increases and as a result, the number of sources that can be connected to the base station decreases, the number of unmatched sources increases in an upward trend. This process continues until at the two endpoints of the chart, where only one source can connect to the base station, the number of unmatched sources reaches 119.

Figure~\ref{fig:s4Unmatched} shows the trend chart of the average number of unmatched sources for all algorithms (DiTOSA, SORSA\_W, SORSA\_B, SORSA\_Z,	DORSA\_W-B, DORSA\_W-Z, DORSA\_B-Z, and DORSA\_W-B-Z) vs. the number of sources in Scenario 4.


			\begin{figure}[!htb]
			\centering
			\includegraphics[scale=0.5]{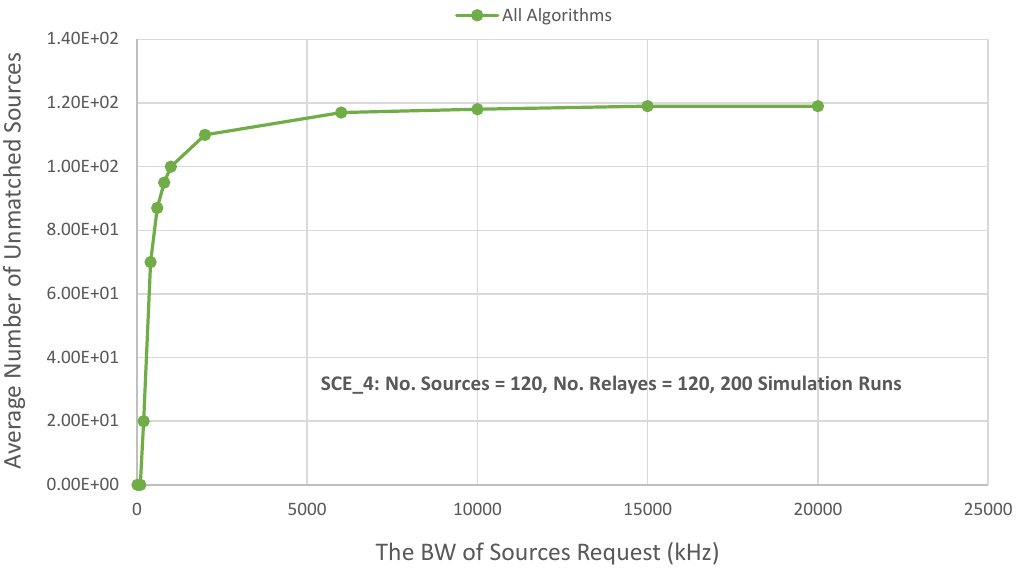}
			\caption[The average number of unmatched sources for all algorithms vs. the BW of sources request (kHz) in Scenario 4 with 120 sources and relays and 200 simulation runs.]{The average number of unmatched sources for for all algorithms vs. the BW of sources request (kHz) in Scenario 4 with 120 sources and relays and 200 simulation runs.}
			\label{fig:s4Unmatched}
			\end{figure} 

	\item \textbf{Actual Execution Time}

As shown in Figure~\ref{fig:s4time}, the average actual execution time of all types of DORSA is a maximum of 2460 (ms) and a minimum of 192 (ms). Also, the $ln$-based trendline obtained for these graphs will not violate the temporal complexity with the upper limit $O(n^3)$ of DORSA algorithms.



	
		\begin{figure}[!htb]
			\centering
			\includegraphics[scale=0.47]{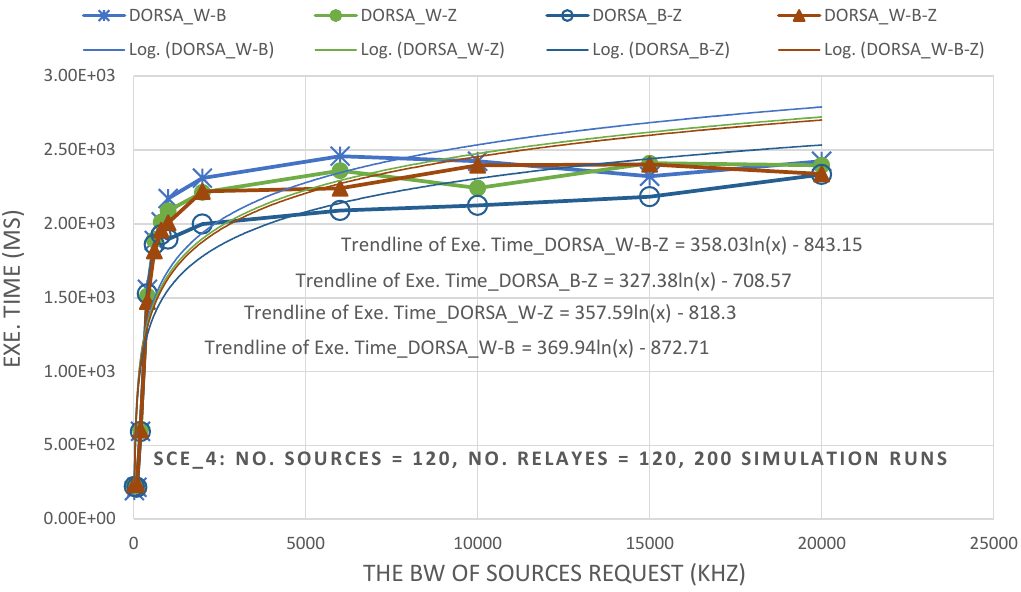}
			\caption[The average actual execution time (ms) and its trendline for DORSA\_W-B, DORSA\_W-Z, DORSA\_B-Z, and DORSA\_W-B-Z vs. the BW of sources request (kHz) in Scenario 4 with 120 sources and relays and 200 simulation runs.]{The average actual execution time (ms) and its trendline for DORSA\_W-B, DORSA\_W-Z, DORSA\_B-Z, and DORSA\_W-B-Z vs. the BW of sources request (kHz) in Scenario 4 with 120 sources and relays and 200 simulation runs.}
			\label{fig:s4time}
		\end{figure}			
	
			\end{enumerate}

\section{Conclusion and Future Work}\label{sec:conclusion}
In this paper, we focus on providing a method for jointly dynamic M2M RF interface setting and relay selection simultaneously. This allows the use of multiple RF interfaces in the network at the same time. Clearly, this in practice can also reduce transmission interference compared to situations where all data transmission is done on a single RF interface in various IoT applications.
%

Thus, the Dynamic Optimal Relay Selection and RF interfaces Setting Algorithm (DORSA) were presented using the Hungarian algorithm. The mean results showed better performance of DORSA along with three RF interfaces in all scenarios compared to other algorithms. DORSA\_W-B-Z performed on average between 0.8-10\% in different scenarios better than other algorithms. The impact of using DORSA was most evident in the scenario where we saw a change in the number of machines, whether source or relay.




In the following, by extending the mentioned method, by selecting the optimal M2B RF interfaces, we will seek to provide a suitable framework for the optimal solution of the relay selection in M2M communications by dynamically setting the RF interfaces.


%
%

\section*{\textbf{Appendix: The Proof of Optimality of DORSA} }\label{subsec:ProofDORSA}
%




In this section, the optimality of the proposed solution for our desired dynamic optimal relay selection and RF interfaces setting (main $k$-AP) is briefly presented. For this purpose, the correspondence between the main $k$-AP and the transformed standard assignment problem is examined.



It has already been proved that the solution provided by the Hungarian algorithm is optimal for the desired standard assignment problem \cite{KuhaHMAP1955}. Now, if it is proved that the main $k$-AP corresponds to the new transformed standard assignment problem, the optimal solution of the new transformed problem by the Hungarian algorithm will be the same as the optimal solution of main $k$-AP. Theorem \ref{th:kAPCorrespondsAP}, similar to Theorem 1 in our previous work \cite{MRSRGhHe2020}, proves that the solutions of main $k$-AP and the desired standard assignment problem correspond to each other.



We know that the main $k$-AP can be formulated in the form of equation~(\ref{eq:maxEdge}).

\begin{align}
\label{eq:maxEdge}
    &\Max_{|SE|=k}
    \begin{aligned}[t]
       &\sum_{e_{i,j} \in  SE}{ w_{e_{i, j}} }  , \\
    \end{aligned}
\end{align}
where $SE$ is the set of selected edges of the main $k$-AP \cite{MRSRGhHe2020}.



Then, due to Step 1 of the solution (in subsection \ref{subsec:solveJRRSS}), a number of new vertices are added to both sides of the graph, with weighted edges with $A_{value}$ to connect the previous nodes and zero weight edges to connect the new nodes on the opposite side.


According to Lemma 1 in subsection III.A.3 of our previous work \cite{MRSRGhHe2020}, it is proved that when transforming the main $k$-AP to a new standard assignment problem, the number of new edges added to the graph will be constant and equal to $n_{A_{SE}} = (n-k)+(m-k) = n+m-2k$.


Now, if $SE$ is the set of selected edges of the new transformed standard assignment problem and $E_{A}$ is the set of $A_{value}$-weighted edges, the mathematical form of the new optimization problem related to the transformed problem can be formulated as equation~(\ref{eq:new_maxEdge_with_Constant_nA}).

\begin{align}
\label{eq:new_maxEdge_with_Constant_nA}
    &\Max_{| \lbrace SE   -   E_{A} \rbrace |= k}
    \begin{aligned}[t]
      &  \lbrace \sum_{{e_{i,j}} \in \lbrace SE - E_{A} \rbrace }  w_{e_{i, j}} +  n_{A_{SE}} \times A_{value} \rbrace , \\
    \end{aligned}  
\end{align}

where as mentioned $n_{A_{SE}}= (n-k)+(m-k)$ is the number of selected $A_{value}$-weighted edges \cite{MRSRGhHe2020}.

Now, according to the points mentioned, it is proved by Theorem \ref{th:kAPCorrespondsAP} that the new transformed problem and the main $k$-AP correspond to each other.


Therefore, the solution provided by DORSA for the main $k$-AP provides the optimal solution for our desired dynamic optimal relay selection and RF interfaces setting. 


\begin{theorem} \label{th:kAPCorrespondsAP}
Each optimal solution for the desired transformed standard assignment problem corresponds to an optimal solution for main $k$-AP and vice versa \cite{MRSRGhHe2020}.
\end{theorem}

\textbf{Proof}: 
We know the answer set of equation~(\ref{eq:maxEdge}) ($S_{(k-AP)}$) has a maximum $k$ edge and the answer set of equation~(\ref{eq:new_maxEdge_with_Constant_nA}) ($S_{(new)}$) has a maximum $m+n-k$ edge. Now, to prove this theorem, it suffices to prove that the set $S_{(k-AP)}$) corresponds to the set $S_{(new)}$ and vice versa. Now,
\begin{itemize}

\item[-] If in each optimal solution for the desired transformed standard assignment problem all $n_{A_{SE}}$ edges with weight $A_{value}$ are removed, the edges between the nodes of the main $k$-AP are an optimal solution for the main $k$-AP. Thus, the new solution set will have $k‌$ assigned edges.



\item[-] If in each optimal solution for the main $k$-AP add an edge with weight $A_{value}$ from the new non-duplicate additional nodes to the unassigned nodes of the main $k$-AP, the new solution will have $m+n-k‌$ edges. This new solution corresponds to an optimal solution for the desired transformed standard assignment problem.



\end{itemize}

  \section*{Acknowledgment}
The authors would like to thank Mr. Mostafa Mahdieh, Mr. Kian Mirjalali, Dr. Hossein Ajorloo, and Dr. Siavash Bayat for their helpful suggestions and guidance.

\begin{IEEEbiography}[{\includegraphics[width=1in,height=1.25in,clip,keepaspectratio]{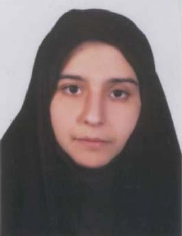}}]{Monireh Allah Gholi Ghasri}
received B.Sc. degree from Amirkabir University of Technology, Tehran, Iran, in 2012 and M.Sc. degree from Sharif University of Technology, Tehran, Iran, in 2014, both in Information Technology (IT) Engineering. Currently, she is working toward the Ph.D. degree in Computer Engineering in Wireless Networking Lab, Sharif University of Technology, Tehran, Iran. Her research interests include wireless networks, Machine-to-Machine (M2M) communications, Internet of Things (IoT), next-generation network management, relay selection, network protocols and architectures, game theory, and matching theory.
\end{IEEEbiography}

\begin{IEEEbiography}[{\includegraphics[width=1in,height=1.25in,clip,keepaspectratio]{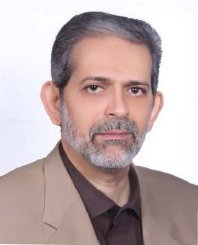}}]{Ali Mohammad Afshin Hemmatyar}
received B.Sc., M.Sc. and Ph.D. degrees in electrical engineering from Sharif University of Technology, Tehran, Iran in 1988, 1991, and 2007, respectively. Since 1991, he has joined Department of Computer Engineering in Sharif University of Technology, where he is currently an assistant professor. His research interests are Vehicular Ad-hoc Networks, Mobile Ad-hoc Networks, Cognitive Radio Networks, Wireless Sensor Networks, and Internet of Things (IoT).
\end{IEEEbiography}

\EOD

\end{document}